\documentclass[a4paper,fleqn,usenatbib]{mnras}

\usepackage{newtxtext,newtxmath}
\usepackage[T1]{fontenc}
\usepackage{ae,aecompl}

\usepackage{soul}
\usepackage{xcolor}

\definecolor{kellygreen}{rgb}{0.3, 0.73, 0.09}

\usepackage{graphicx}	% Including figure files
\usepackage{amsmath}	% Advanced maths commands
\usepackage{amssymb}	% Extra maths symbols

\title[Massive discs around low-mass stars]{Massive discs around low-mass stars}
\author[T. J. Haworth et al.]
{\parbox{\textwidth}{Thomas J. Haworth$^{1, 2}$\thanks{E-mail: \texttt{t.haworth@imperial.ac.uk}},
James Cadman$^{3,4}$ 
Farzana Meru$^{5,6}$,
Cassandra Hall$^{7,8}$, \\
Emma Albertini$^{2}$, 
Duncan Forgan$^9$,
Ken Rice$^{3,4}$ and James E. Owen$^2$
}\vspace{0.4cm}\\
\parbox{\textwidth}{$^{1}$ Astronomy Unit, School of Physics and Astronomy, Queen Mary University of London, London E1 4NS, UK \\ $^{2}$ Astrophysics Group, Imperial College London, Blackett Laboratory, Prince Consort Road, London SW7 2AZ, UK\\
$^{3}$ SUPA, Institute for Astronomy, University of Edinburgh, Royal Observatory, Blackford Hill, Edinburgh, EH93HJ, UK \\
$^4$ Centre for Exoplanet Science, University of Edinburgh, Edinburgh, UK \\
$^{5}$ Centre for Exoplanets and Habitability, University of Warwick, Gibbet Hill Road, Coventry CV4 7AL, UK\\
$^{6}$ Department of Physics, University of Warwick, Gibbet Hill Road, Coventry CV4 7AL, UK\\
$^{7}$ Dept. of Physics \& Astronomy, University of Leicester, University Road, Leicester, LE1 7RH, UK \\
$^{8}$ School of Physics and Astronomy, Monash University, VIC 3800, Australia\\
$^{9}$ Centre for Exoplanet Science, SUPA, School of Physics \& Astronomy, University of St Andrews, St Andrews KY16 9SS, UK }}

\date{Accepted XXX. Received YYY; in original form ZZZ}

\pubyear{2020}

\begin{document}
\label{firstpage}
\pagerange{\pageref{firstpage}--\pageref{lastpage}}
\maketitle

\begin{abstract}

We use a suite of SPH simulations to investigate the susceptibility of protoplanetary discs to the effects of self-gravity as a function of star-disc properties. We also include passive irradiation from the host star using different models for the stellar luminosities. The critical disc-to-star mass ratio for axisymmetry (for which we produce criteria) increases significantly for low-mass stars. This could have important consequences for increasing the potential mass reservoir in a proto Trappist-1 system, since even the efficient Ormel et al. (2017) formation model will be influenced by processes like external photoevaporation, which can rapidly and dramatically deplete the dust reservoir. The aforementioned scaling of the critical $M_d/M_*$ for axisymmetry occurs in part because the Toomre $Q$ parameter has a linear dependence on surface density (which promotes instability) and only an $M_*^{1/2}$ dependence on shear (which reduces instability), {but also occurs because, for a given $M_d/M_*$, the thermal evolution depends on the host star mass}. The early phase stellar irradiation of the disc (for which the luminosity is much higher than at the zero age main sequence, particularly at low stellar masses) can also play a key role in significantly reducing the role of self-gravity, meaning that even Solar mass stars could support axisymmetric discs a factor two higher in mass than usually considered possible. We apply our criteria to the DSHARP discs with spirals, finding that {self-gravity can explain the observed spirals so long as the discs are optically thick to the host star irradiation.}

\end{abstract}


%%%%%%%%%%%%%%%%%%%%%%%%%%%%%%%%%%%%%%%%%%%%%%%%%%

%%%%%%%%%%%%%%%%% BODY OF PAPER %%%%%%%%%%%%%%%%%%

%\newpage

\section{Introduction}

The gravitational instability (GI) of circumstellar discs has important consequences for the transport of angular momentum, placing constraints on the possible mass/radius of a disc and ultimately for understanding planet formation \citep{1987MNRAS.225..607L, 2010MNRAS.402.1740R}. At some mass, self-gravity will cause deviations from axisymmetry through spirals which can survive in a quasi-stable state for extended periods of time \citep{2001ApJ...553..174G, 2004MNRAS.351..630L}. If a disc cools quickly enough relative to its dynamical timescale, it may even fragment \citep{2001ApJ...553..174G,2003MNRAS.339.1025R}, ultimately forming stars \citep{2016Natur.538..483T}, substellar companions and giant planets \citep{2013MNRAS.432.3168F,2017MNRAS.470.2517H} , and, more rarely, terrestrial mass planets through tidal downsizing (see, e.g. \citealt{nayakshin2010,forganetal2018}).

Many aspects of GI are now very well known \citep[for a review see][]{2016ARA&A..54..271K}. In a rotating disc, gravitational collapse has to overcome shear in addition to thermal pressure. This competition is encapsulated by the Toomre Q parameter
\begin{equation}
    Q = \frac{c_s\Omega}{\pi G \Sigma},
    \label{equn:simpleQ}
\end{equation}
with discs eventually becoming { gravitationally} unstable for $Q<1-2$. Here $c_s$ is the sound speed, $\Sigma$ is the surface density and $\Omega$ is the epicyclic frequency.  {For a given stellar mass, we would expect $Q$ to decrease as the disc mass increases, if the sound speed remains constant. Hence, there is a disc mass above which we'd expect the disc to become gravitationally unstable.}  {For discs still accreting from some natal core, the disc mass will increase if the mass accretion rate through the disc is smaller than the ambient accretion rate onto the disc.  We would therefore expect such discs to eventually become gravitationally unstable \citep{2010ApJ...708.1585K, 2011MNRAS.417.1839H}.}

{ One way to estimate the gravitational stability of a disc is to consider the disc-to-star mass ratio.}  For example, \cite{2016ARA&A..54..271K} re-write equation \ref{equn:simpleQ} in terms of this ratio, assuming $Q<\sim 1$ for instability, to give 
\begin{equation}
    \frac{M_d}{M_*} > 0.06 f  \left(\frac{T}{10\,K}\right)^{1/2}\left(\frac{R}{100\,\textrm{AU}}\right)^{1/2}\left(\frac{M_*}{M_\odot}\right)^{-1/2}
    \label{equn:KratterLodato}
\end{equation}
as the criterion for instability, where $f$ subsumes factors of order unity related to the surface density profile. If the disc temperature term doesn't vary much with stellar mass, then this expression implies that low-mass stars should be able to support relatively high disc-to-star mass ratios. The temperature of inner parts of the disc \textit{will} certainly vary with stellar mass, but at the larger radii where gravitational instability operates the temperature will {probably} have reached some floor value $T_{\rm floor} \sim 10$\,K. 

Higher stable $M_d/M_*$ for low-mass stars could be extremely important for explaining the abundance of planets around low-mass stars such as Trappist-1, where the planet formation efficiency would have to be extremely high if ${M_d}/{M_*}$ were the canonical $\sim0.1$ ({the typical value of the disc aspect ratio, $H/R$}). This was highlighted by \cite{2018MNRAS.475.5460H}, who studied the evolution of discs around possible Trappist-1 precursors including accretion of dust grains onto the parent star and entrainment of dust in a photoevaporative wind due to irradiation by nearby stars in the natal stellar cluster. These models demonstrated that $M_d/M_*$ has to be $>0.1$ to retain enough solids to produce the $\sim5.5$\,M$_{\oplus}$ of planets observed towards that system \textit{so far} \citep{2017Natur.542..456G, 2017arXiv170404290W}, even if extremely high planet formation efficiencies were permitted \citep[e.g.][]{2017A&A...604A...1O,2019arXiv190600669S}. According to \cite{2018MNRAS.475.5460H}, for $M_d/M_*=0.1$ it is often rendered \textit{impossible} to produce the planets we observe in Trappist-1, as the mass reservoir drops too rapidly ($<<1$\,Myr). Conversely, if the initial disc-to-star mass ratio for a proto-Trappist-1 were more massive than canonically expected then { GI should not lead to so widespread fragmentation \citep[consistent with][]{2011ApJ...731...74B} since the masses of the planets we are detecting are too low to have resulted from that mechanism \citep{2011MNRAS.418.1356R}. The higher disc mass would either have to remain stable, or sustain a quasi-steady state in which spiral density waves may develop but in which fragmentation does not happen \citep{2004MNRAS.351..630L}.} 

Another challenge for planet formation around low-mass stars is that the radial drift of dust grains occurs at higher velocity due to the lower angular velocity at any given radius \citep{1977MNRAS.180...57W, 1986Icar...67..375N, 2006ApJ...636.1121J, 2012A&A...539A.148B}. This leads to more rapid accretion onto the star and greater likelihood of destructive collisions. A larger disc mass reservoir would help to address the problem this poses to planet formation {and if spirals can be present without fragmenting, they may accelerate planetesimal growth to then potentially form planets via Core Accretion \citep{2004MNRAS.355..543R}.}

{Finally, \cite{2019arXiv190912174M} recently discovered a 0.46 Jupiter mass planet around a 0.12\,M$_\odot$ star on a 204\,day orbit, which they propose formed via GI, further suggesting that massive discs may exists around low-mass stars. }

Despite the potential importance of higher disc-to-star mass ratios at low stellar masses, it is a commonplace assumption that the maximum value is $\sim 0.1$. {However it is known that higher disc-to-star mass ratios can be sustained with self-gravitating discs that are quasi-stable \citep[e.g.][]{2011MNRAS.410..994F}. Nevertheless} it is not an unreasonable approach to have taken; even given the expected $M_*^{-1/2}$ scaling, an assumed {upper limit on the} ratio of $\sim 0.1$ should be accurate to within about 50\,per cent down to the upper limit on M dwarf masses. However the stable disc-to-star mass ratio could deviate by a, potentially critical, factor $\sim3.5$ by the 0.08\,M$_\odot$ stellar mass of Trappist-1. This overlooked scaling could be very important for understanding, apparently very prolific, planet formation in the low stellar mass regime. The potential importance of higher disc-to-star mass ratios, permitting quasi-steady self-gravitating discs was recently highlighted by \cite{2018MNRAS.477.3273N} in their discussion of the maximum mass Solar nebula. They argue that this closer link to the initial properties as determined by the star formation process, including features like spirals that concentrate dust, can play a key role in the the early production of planetesimals.

The scaling of the maximum stable disc mass with stellar mass and disc size in equation \ref{equn:KratterLodato} is also relatively simple. It is the variation of the disc temperature which is most difficult to gauge. Not only is the temperature an important quantity for determining the onset of instability, it also sets the nature of the subsequent instability (such as how it fragments). An approach to considering the thermal properties, without having to include complex microphysics, has involved parameterising the ratio of the cooling timescale to the dynamical timescale using $\beta\equiv 2\pi t_{\rm cool}/t_{\rm dyn}$. For large enough $\beta$ (typically $>\sim3$), cooling is slower than heating and the disc {will tend to be} stable against fragmentation \citep{2001ApJ...553..174G, 2005MNRAS.364L..56R}. Due to computational reasons, calculations setting this $\beta$ using a variety of prescriptions have been extremely popular for investigating GI \citep[e.g.][]{2005ApJ...619.1098M, 2011MNRAS.411L...1M, 2012MNRAS.427.2022M}, but miss the details of the actual thermodynamics.  Other more direct cooling treatments have been implemented, such as the approximate radiative transfer schemes of \cite{2004MNRAS.353.1078W}, \cite{2007A&A...475...37S}, \cite{2009MNRAS.394..882F}  and \cite{2015MNRAS.447...25L}. A comparison of some of these methods, including  against $\beta$ cooling models, is made by \cite{2018MNRAS.478.3478M}.  In addition to these more sophisticated cooling methods, radiation from the central star can also increase the temperature, which is not accounted for in $\beta$ cooling models, but has been shown to generally stabilise the disc against fragmentation \citep[e.g.][]{2008ApJ...673.1138C, 2010MNRAS.406.2279M, 2011ApJ...740....1K, 2011MNRAS.418.1356R, 2016MNRAS.458..306H}

From \cite{2016ARA&A..54..271K} and equations \ref{equn:simpleQ} and \ref{equn:KratterLodato} we do have an anticipated scaling of disc stability with stellar mass. However this has not yet been directly confirmed with simulations {\citep[][model isothermal discs around 0.3\,$M_\odot$ M stars for discs of radius from 1/3 to 30\,AU and for $M_d/M_*$ from 0.033 to 0.266 and find GI can occur, but do not model how the behaviour is sensitive to stellar mass]{2016MNRAS.463.2480B}}. {An additional issue is that the evolution of a self-gravitating disc becomes increasingly global for disc-to-star mass ratios above 0.5 \citep{2011MNRAS.410..994F}. Hence, some of the simple scalings no longer apply.} Though note that \cite{2005MNRAS.358.1489L} found that even massive discs seem to roughly obey the fragmentation criteria, requiring cooling on shorter than dynamical timescales. 

In this paper we present a suite of {3D SPH simulations} of discs of different mass and extent for different stellar masses. We also compute models with and without irradiation of the disc by the host star. Our aim is to determine for what disc-to-star mass ratios discs remain axisymmetric, as well as the {qualitative} nature of any spirals {and whether or not such discs are likely to undergo fragmentation}, as a function of stellar mass. In particular we aim to investigate whether systems such as Trappist-1 can sustain large disc-to-star mass ratios {without fragmenting} and, like \cite{2018MNRAS.477.3273N}, highlight that it should not be taken for granted that the upper limit on the disc mass is always just 0.1\,M$_*$.

\begin{figure*}
    \centering
    \vspace{-0.5cm}
    \includegraphics[width=8.4cm]{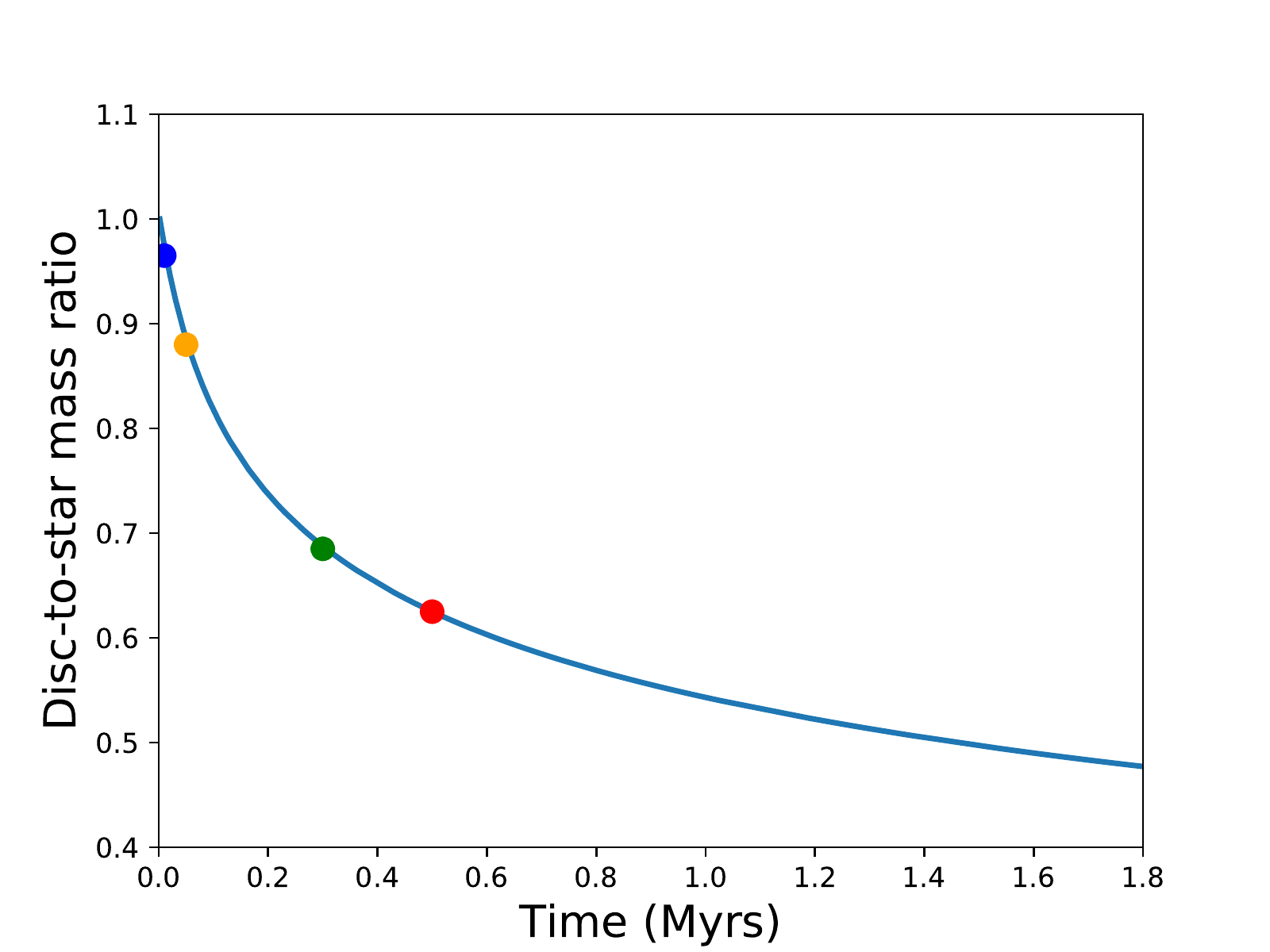}
    \includegraphics[width=8.4cm]{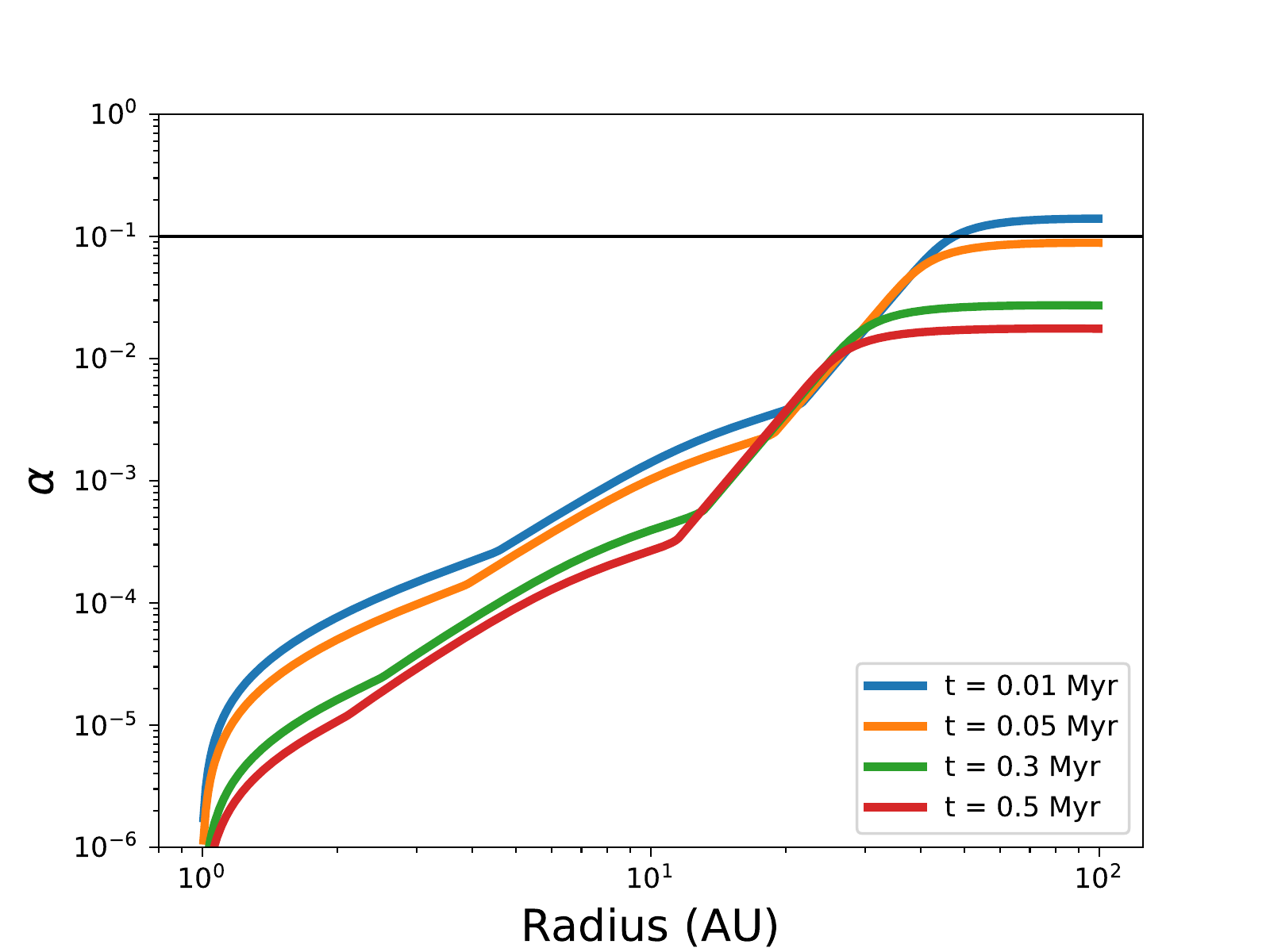}
    
    \vspace{-0cm}
    \includegraphics[width=8.4cm]{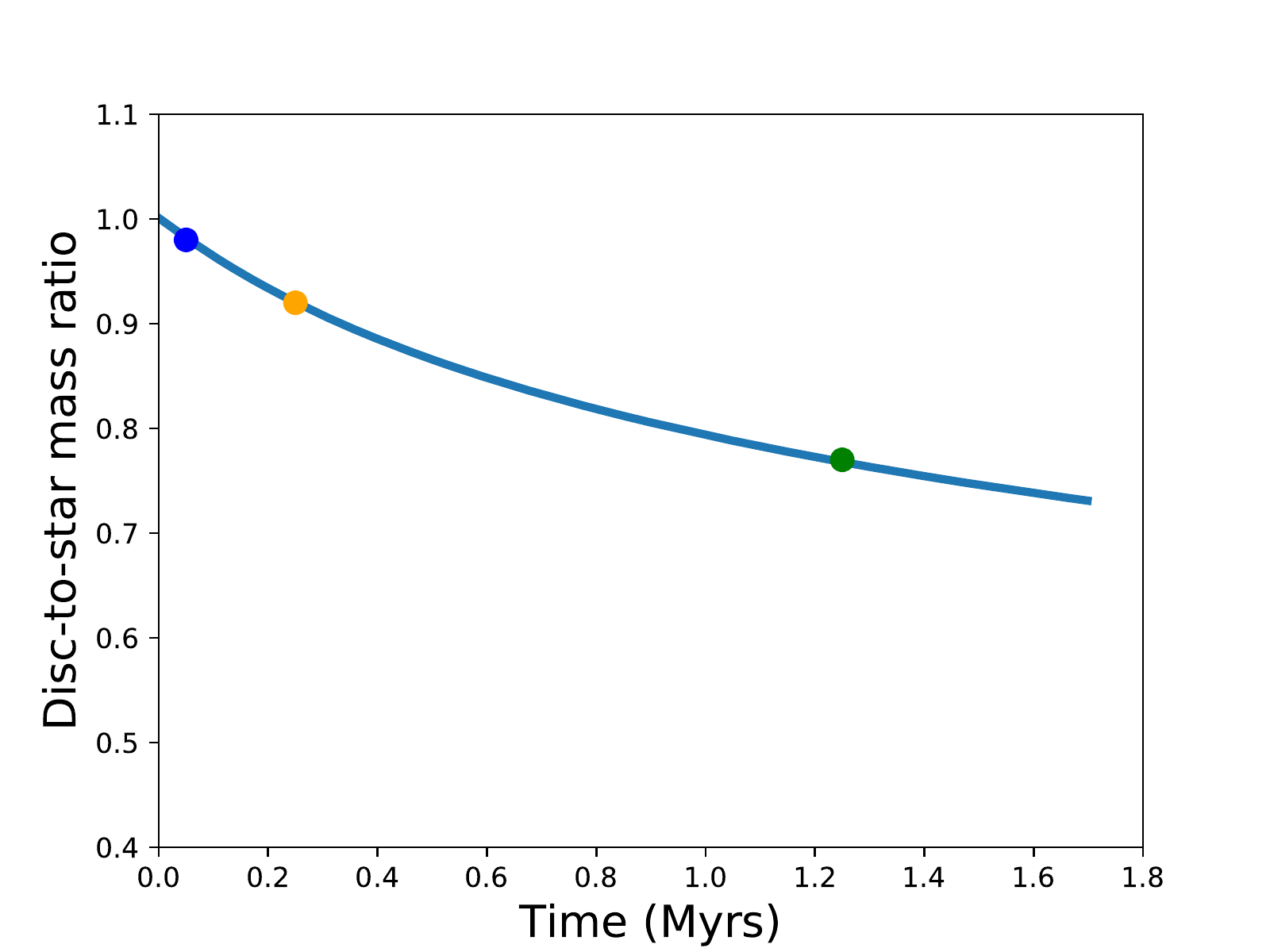}
    \includegraphics[width=8.4cm]{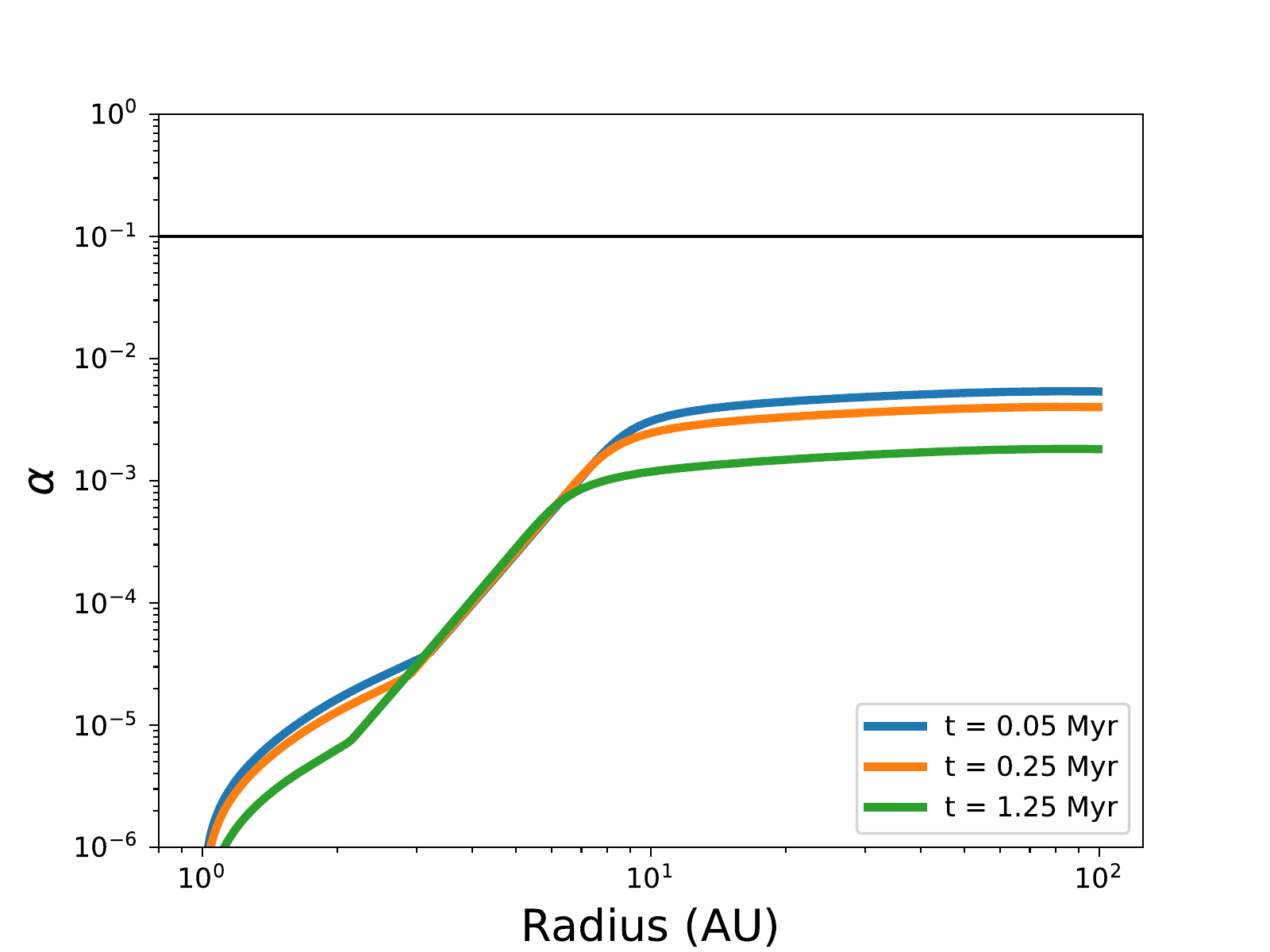}    
    \caption{Time evolution of 1D self-gravitating discs for a Solar mass star (upper panels) and 0.1\,M$_\odot$ star (lower panels). The left hand panel is the disc-to-star mass ratio over time and the right hand panel is the radial effective viscous $\alpha$ profile at different snapshots in time, demarked by the  points in the left hand panel. The horizontal black line in the right hand panels represents $\alpha=0.1$, the approximate value above which fragmentation is expcted.  }
    \label{fig:1Devo}
\end{figure*}

\section{Prologue: Expectations from 1D models}
The main focus of this paper is (radiation) hydrodynamic calculations of discs to probe the gravitational stability for different star/disc parameters. However we begin by directing existing semi-analytic models to the problem to quickly gauge the expected impact on stability of varying the stellar mass. {Note that these illustrative 1D models assume that angular momentum and energy transport via gravitational instability can be regarded as a local phenomenona \citep{2004MNRAS.351..630L}, which breaks down for disc-to-star mass ratios $ M_d/M_* > 0.5$ \citep{2011MNRAS.410..994F}. We do, however, apply it to situations where there are disc-to-star mass ratios $M_d/M_* > 0.5$, in which case global effects - that are not considered - do become important. Hence, these 1D models should simply be treated as extrapolations from existing semi-analytic models to further motivate the SPH simulations that follow.}

\subsection{1D Time dependent self-gravitating disc models}
\label{sec:timedep1D}

Before delving into full hydrodynamic simulations,  we consider the longer term evolution of self-gravitating discs around stars of different masses. We do so using the 1D framework of \cite{2009MNRAS.396.2228R}. It is described in full in that paper and also neatly summarized by  \cite{2019ApJ...871..228H}, so we do not describe the algorithm in any detail here. Though we note that it involves solving the \cite{1974MNRAS.168..603L} surface density evolution equation assuming thermal equilibrium and angular momentum transport dominated by self-gravity. {We include the gravitational potential both from the host star and the disc itself.} These models do not account for heating due to irradiation by the host star. Although these calculations are not able to capture the full nature of stability of the discs, the effective $\alpha$ viscosity does give soft limits on the behaviour.  { The general expectation is that if the pseudo-viscous $\alpha$ is less than $\sim 0.1$, the disc can settle into a quasi-steady state in which GI acts to transport angular momentum outwards, allowing mass to accrete onto the central protostar.  On the other hand, if  the pseudo-viscous $\alpha$ exeeds $\sim 0.1$, a disc is likely to fragment to form bound objects, potentially of planetary mass \citep{2001ApJ...553..174G, 2003MNRAS.339.1025R, 2005MNRAS.364L..56R}}. 

We computed the time evolution of discs around 0.1 and 1.0\,M$_\odot$ stars, with initial disc-to-star mass ratios of $\sim1$ and an outer radius of 100\,AU. This initial disc mass is high, but such values do appear in the simulations of star formation that naturally produce discs \citep{2018MNRAS.475.5618B}. The evolution of the disc-to-star mass ratio and  self-gravitationally driven effective $\alpha$ viscosity is given for each model in Figure \ref{fig:1Devo}. 

In the Solar mass case (the upper panels of Figure \ref{fig:1Devo}) the disc mass depletes rapidly, dropping {40\,per cent} by 0.5\,Myr. This is due to the high effective $\alpha$ facilitating rapid viscous evolution, particularly beyond about 20\,AU as illustrated in the upper right hand panel of Figure \ref{fig:1Devo}. The effective viscosity, $\alpha$, is indeed high enough that disc fragmentation would be expected to occur beyond $r\sim30$\,AU, and such conditions could persist for tens of kyr. 

In the 0.1\,M$_\odot$ case (lower panels) the fractional depletion of the disc-to-star mass ratio is less rapid, which is set by the correspondingly smaller $\alpha$ profile in the lower right-hand panel. In addition, the disc is never expected to fragment, even very early on. Low-mass stars would therefore be expected to host high disc-to-star mass ratios for million year timescales without fragmenting.

This smaller effective $\alpha$ in the lower stellar mass case arises because when we have: 
\begin{itemize}
    \item $Q\sim1$
    \item energy balance
    \item the same disc size
    \item the same disc-to-star mass ratio
\end{itemize} 
the cooling (and hence heating) rates are smaller in the lower stellar mass case. The effective $\alpha$ is directly proportional to this cooling rate \citep[see][]{2009MNRAS.396.2228R}

So the expectation from these existing 1D evolutionary models is indeed that lower mass stars can support higher disc-to-star mass ratios {without them undergoing fragmentation} that, due to the associated lower effective viscosity, can also be retained over long periods of time.

\subsection{Constraints from self-gravitating, pseudo-viscous accretion disc models}
\label{sec:1Dacc}
\begin{figure}
    \centering
    \includegraphics[width=9.4cm]{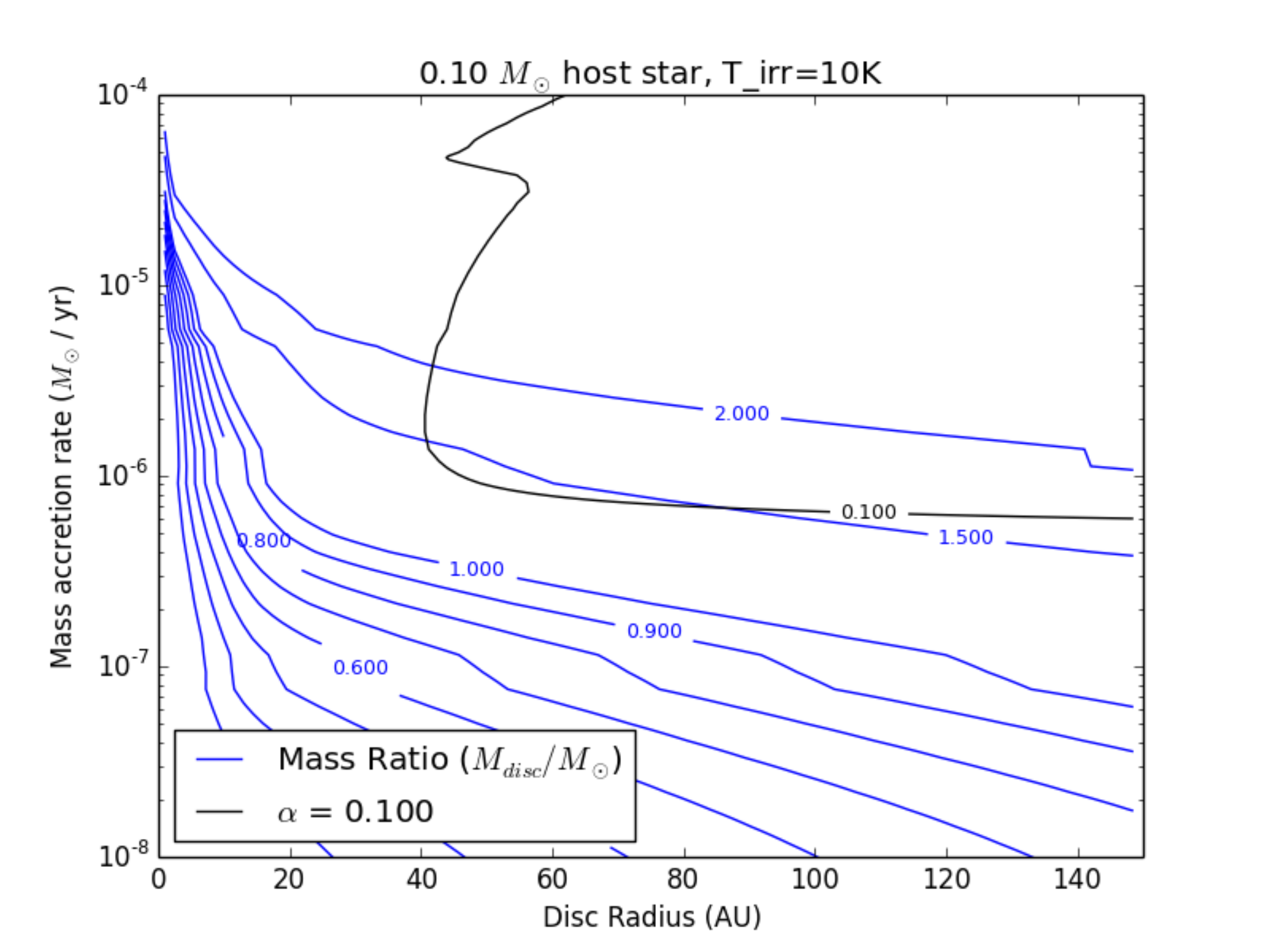} 
    \includegraphics[width=9.4cm]{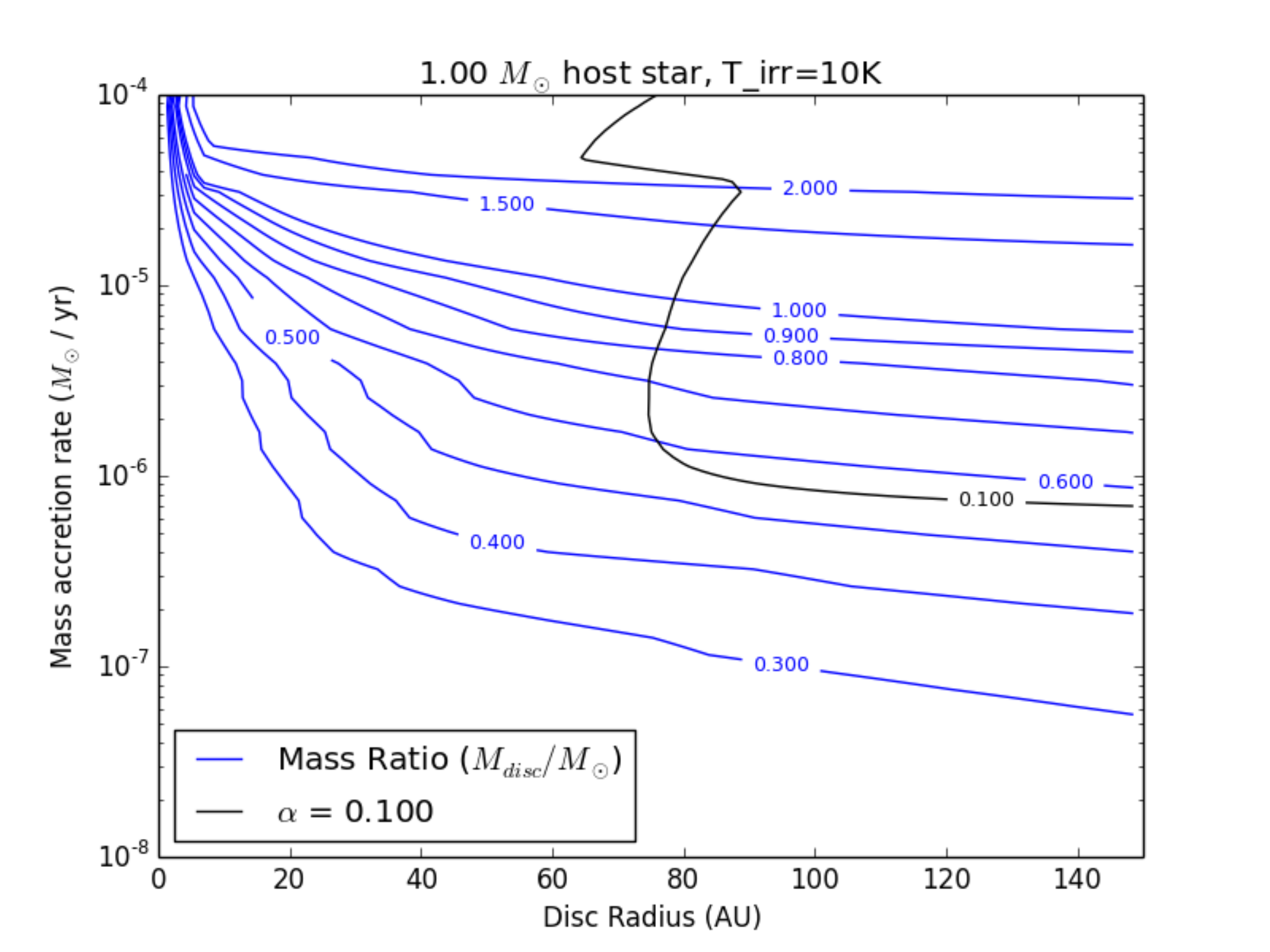}    
    \caption{The mass accretion rate as a function of disc radius for different contours of disc-to-star mass ratio (blue). The upper and lower panels are for 0.1\,M$_\odot$ and 1\,M$_\odot$ stars respectively. The black contour denotes the $\alpha=0.1$ boundary (alpha increases towards the upper right) which approximately separates the regions of fragmenting and non-fragmenting parameter space. In the 0.1\,M$_\odot$ case only disc-to-star mass ratios greater than $\sim1.5$ are expected to fragment, and even then only beyond around 40AU. In the 1\,M$_\odot$ the fragmentation disc-to-star mass ratio is much lower. }
    \label{fig:RiceMdot}
\end{figure}

We also follow \cite{2011MNRAS.417.1928F} and implement their rapid approach to computing the disc size and disc-to-star mass ratio for a given accretion rate. This approach is based on that of \cite{2009MNRAS.396.1066C}, \cite{2009MNRAS.396.2228R} \citep[see also][]{2015ApJ...804...62R}. We set up a $Q=2$ disc at all radii, with a steady accretion rate and $\alpha$ set by local thermodynamic equilibrium \citep{2010MNRAS.401.2587C}. We assume a floor temperature of $T_{\rm floor} = 10$\,K in these models. This is shown for the case of a 0.1 and 1\,M$_\odot$ star in the upper and lower panels of Figure \ref{fig:RiceMdot} respectively. The blue contours in this figure denote different disc-to-star mass ratios for different accretion rates as a function of disc size. So for example, in the 0.1\,M$_\odot$ case a disc with accretion rate of $10^{-7}$\,M$_\odot$\,yr$^{-1}$ and an outer radius of 50\,AU would  have a disc-to-star mass ratio of $M_d/M_* \sim 0.7$. If the disc were 100\,AU for the same accretion rate the mass ratio would increase to $M_d/M_* \sim 0.9$, and so on. { Again, we are considering disc-to-star mass ratios that are high enough that the local approximation will no longer strictly apply. However, according to \cite{2004MNRAS.351..630L} and \cite{2011MNRAS.410..994F} this approximation is reasonable for mass ratios $M_d/M_* < 0.5$ and still roughly applies for higher mass ratios, especially if the key factor of interest is whether or not fragmentation is likely to occur. }

The black contours in Figure \ref{fig:RiceMdot} denotes a pseudo-viscous $\alpha$ of $\alpha = 0.1$. To the upper right of that contour is the region of parameter space where we would expect fragmentation to occur. Overall then, these results also point towards stable high disc-to-star mass ratios for low-mass stars. In the 0.1\,M$_\odot$ case, discs are only susceptible to fragmentation for mass ratios $>1.4$ and even then only beyond 40\,AU. In the Solar mass case discs can fragment for disc-to-star mass ratios above $M_d/M_* \sim 0.6$. This value of 0.6 is higher than the canonical $M_d/M_* \sim 0.1$ and also higher than the results from the similar prior models of \cite{2009MNRAS.396.1066C}. The difference with the latter is due in part to our imposing a floor temperature \citep[none is imposed in ][]{2009MNRAS.396.1066C} and in part because the cooling function employed here scales with optical depth $\tau$ as $(\tau+1/\tau)^{-1}$ (to account for the cooling becoming inefficient in the optically-thin regime) whereas that of \cite{2009MNRAS.396.1066C} scales as $1/\tau$. In optically thin regions these cooling functions clearly diverge, which results in the differing behaviour (we ran comparison models to confirm this). A full discussion of the differences between these approaches will be forthcoming in \cite{2020MNRAS.492.5041C}. 

So again, these 1D models are consistent with the general {expectation that high disc-to-star mass ratios around low-mass stars can be stable against fragmentation.}  However, although the stability of discs around Solar mass stars is well studied with hydrodynamic simulations, the stability of these possible high disc-to-star mass ratios at low stellar masses actually now needs to be verified with hydrodynamic models, to which we dedicate the rest of this paper. 

%\section{Numerical Method}
\section{SPH models of self-gravitating discs}
We use the smooth particle hydrodynamics code \textsc{sphNG} for the calculations in this paper \citep{1995MNRAS.277..362B}. This code has been applied to a number of gravitationally unstable (and stable) disc applications in the past \citep[e.g. recent works include][]{2012MNRAS.427.2022M, 2015MNRAS.453L..78L, 2018MNRAS.477.4241L, 2019MNRAS.486.2587W}. 

The temperature is of importance to the stability of the disc (see equations \ref{equn:simpleQ}, \ref{equn:KratterLodato}) providing thermal support against collapse.  Cooling is implemented using the hybrid radiative transfer scheme presented by \cite{2009MNRAS.394..882F} ({we are not using a $\beta$ cooling scheme}) also implemented and discussed in \cite{2011MNRAS.410..994F}. This couples the polytropic cooling of \cite{2007A&A...475...37S} with a flux limited diffusion scheme \citep[e.g.][]{2004MNRAS.353.1078W}. We hence use the associated non-trivial equation of state discussed in \cite{2009MNRAS.394..882F}.  \

Heating processes in these models include $\int P\textrm{d}V$ work, shock heating and in most of our models we also account for  passive irradiation by the host star, which through simple energy balance gives a minimum  disc temperature
\begin{equation}
    T_d = \left(\frac{L_*}{4\pi\sigma R^2}\right)^{1/4}
    \label{equn:T_passive}
\end{equation}
at radius $R$, for stellar luminosity $L_*$  \citep{10.1143/PTPS.70.35}. This minimum disc temperature {is in the limit of an optically thin disc}, so in reality the true minimum temperature may be somewhat lower. However, since we include non-irradiated models we also account for the {opposite} limiting case of an extremely optically thick disc. {Determining the temperature accurately by accounting for the disc optical depth is beyond the scope of this work. However, we also use dust radiative equilibrium models to make an initial assessment of how the passive irradiation temperature may compare in section \ref{sec:radeq}. }

\begin{figure}
    \centering
    \includegraphics[width=9cm]{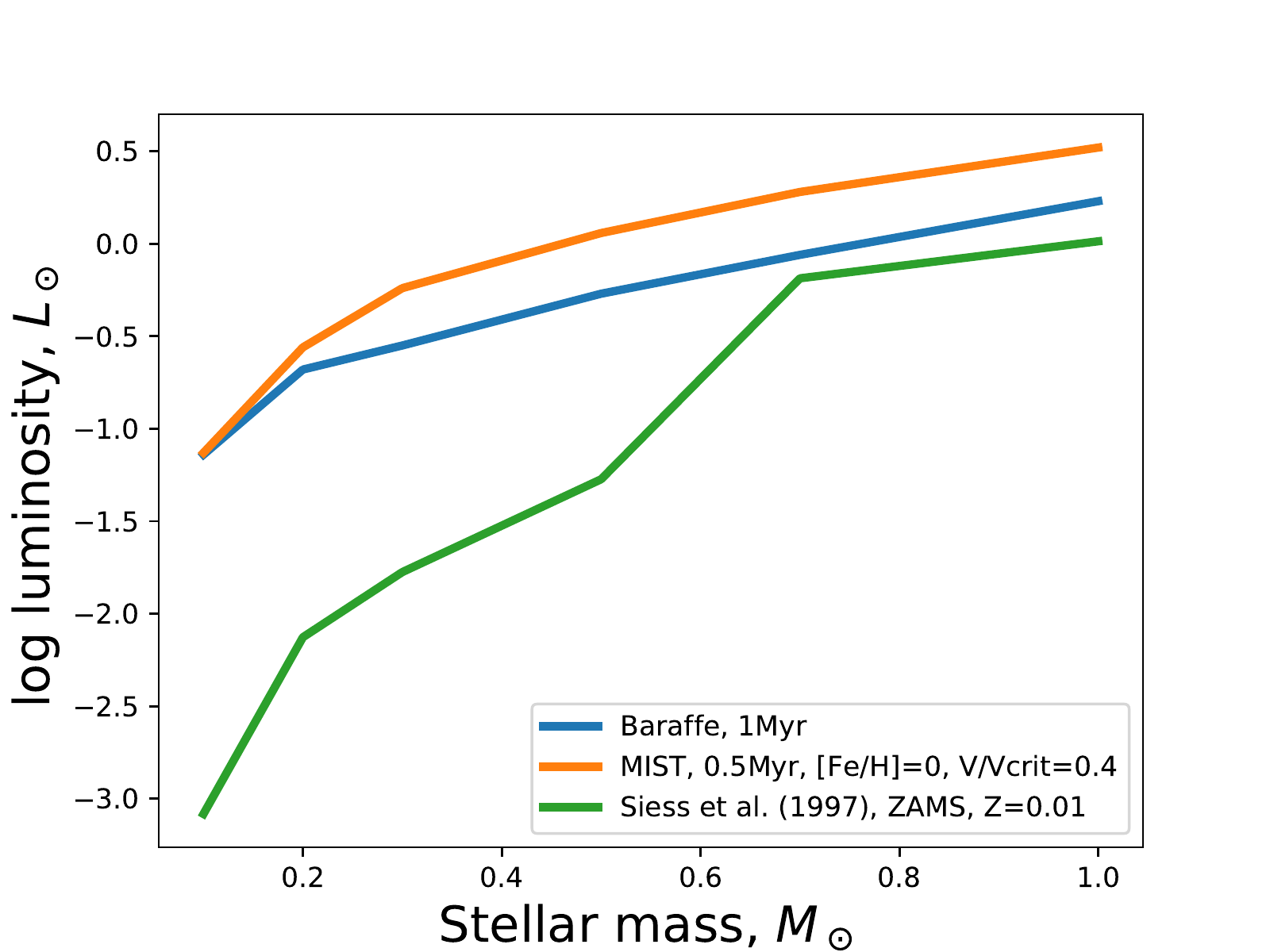}
    \caption{Stellar luminosity as a function of mass for the models in this paper. We run our SPH calculations for both the MIST evolutionary track luminosities at 0.5\,Myr and the \protect\cite{2000A&A...358..593S} ZAMS models. Note that in addition we also run a suite of models with no internal irradiation by the host star. Included are the \protect\cite{1998A&A...337..403B, 2002A&A...382..563B} models at 1\,Myr (the earliest time in those models) for reference  }
    \label{fig:luminosities}
\end{figure}

At early times in the star/disc lifetime when GI is expected to onset, the scaling of stellar luminosity with mass is much flatter than on the main sequence. We therefore estimate luminosities for the passive heating at 0.5\,Myr using MESA Isochrones and Stellar Tracks (MIST\footnote{http://waps.cfa.harvard.edu/MIST/index.html}) evolutionary tracks \citep{2011ApJS..192....3P, 2013ApJS..208....4P, 2015ApJS..220...15P,  2016ApJS..222....8D, 2016ApJ...823..102C}. We use the default initial $v/v_{crit}=0.4$ (rotation velocity to critical/break-up rotation velocity) and a Solar metallicity. The resulting luminosities are similar to those of \cite{1998A&A...337..403B, 2002A&A...382..563B} at 1\,Myr (the earliest available time in those models) as illustrated in Figure \ref{fig:luminosities}, which shows stellar luminosity as a function of stellar mass for these different models. For comparison, we also run a version of the grid using the \cite{2000A&A...358..593S} $Z=0.01$ zero age main sequence (ZAMS) luminosities, which have a much steeper dependence on stellar mass (due to the difference in stellar age, not metallicity) and orders of magnitude lower luminosity for the lowest mass stars. These latter models are also summarised in Figure \ref{fig:luminosities}. Applying these ZAMS luminosities is not really physical at such an early time, however we include models using these luminosities for comparison. We do this is in part because the ZAMS luminosities have been used in other studies of irradiated self-gravitating discs and in part because it serves as a useful guide for the expected behaviour at some intermediate luminosity. However we will not provide criteria for axisymmetry in the ZAMS luminosity case. 

\begin{figure}
    \centering
    \includegraphics[width=9cm]{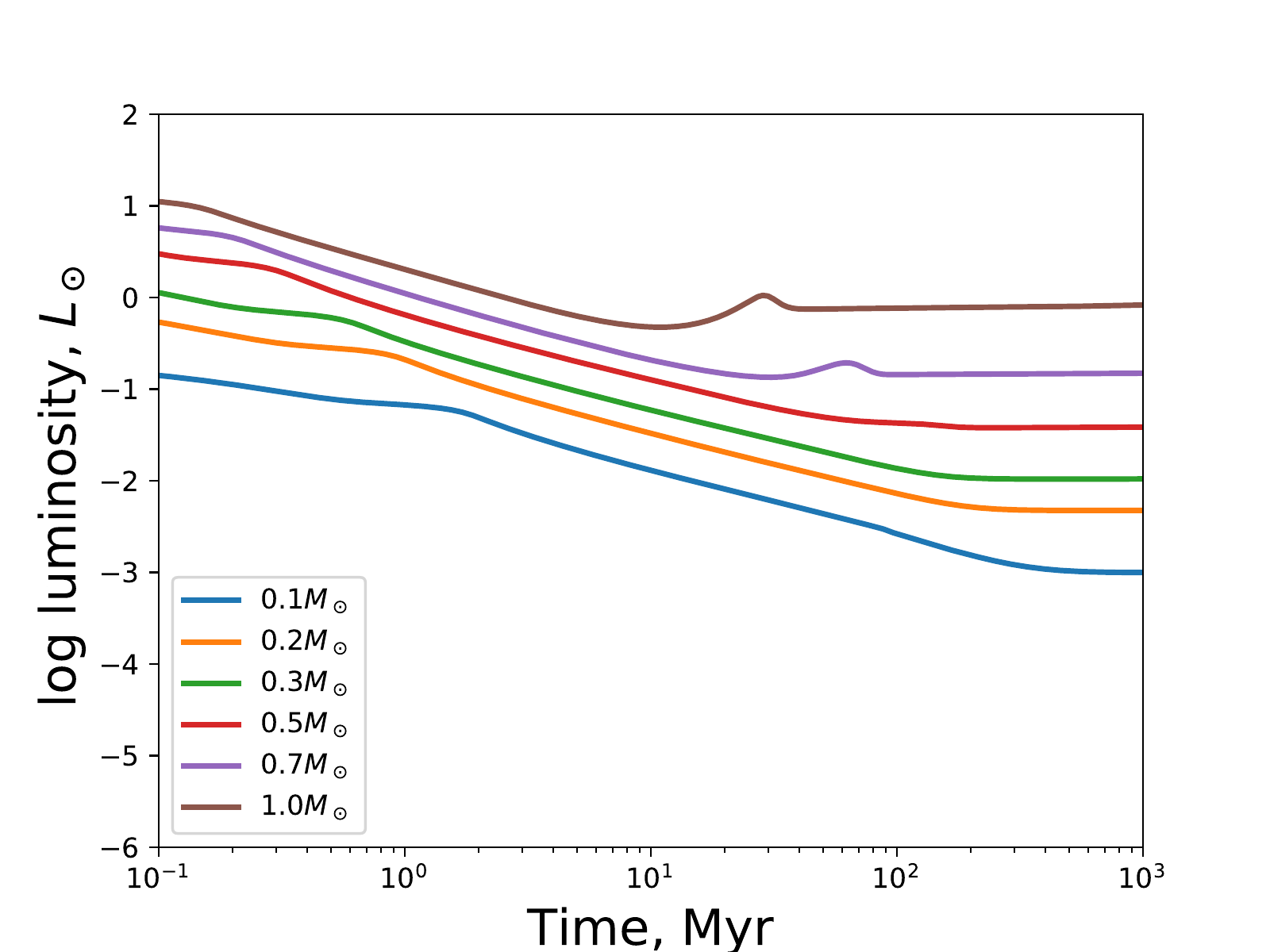}
    \caption{MIST stellar luminosity evolutionary tracks for the stellar masses considered in this paper. See the MIST papers for further details on these tracks   \citep{2011ApJS..192....3P, 2013ApJS..208....4P, 2015ApJS..220...15P,  2016ApJS..222....8D, 2016ApJ...823..102C}}
    \label{fig:luminosityEvo}
\end{figure}

The time evolution of the MIST luminosities is given in Figure \ref{fig:luminosityEvo}. Each of our calculations runs for a maximum of order tens of thousands of years, over which the stellar luminosity will negligibly change (either from the 0.5\,Myr or ZAMS value). For simplicity, we therefore assume a constant stellar luminosity in any given calculation. 

Our treatment of the stellar luminosity is relatively simple and neglects the fact that there will likely be a large column in the very inner regions of the disc. We have not accounted for factors like self-shielding either, which may act to lessen the impact of irradiation in heating the outer disc. However, we note that we also include models with zero irradiation from the host star, which provides the limiting case for additional extinction. In reality it is likely that the actual heating is somewhere between that of our irradiated and non-irradiated models. 

{We use a standard $\alpha-\beta$ {artificial} viscosity scheme with $\alpha_{\rm art}=0.1$ and $\beta=0.2$ \citep[e.g. see section 3.3 of][]{2011MNRAS.410..994F}. As shown in \citep{2011MNRAS.410..994F} this produces, for these kind of simulations, an artificial viscosity with an effective $\alpha_{\rm art}$ of between $\alpha_{\rm art} = 0.001$ and $\alpha_{\rm art} = 0.01$. Although this will act as an additional heating source that will tend to weaken the gravitational instability, these values are well below the $\alpha \sim 0.1$ required for fragmentation. Hence, in self-gravitating discs that are close to the fragmentation boundary, the pseudo-viscosity from the gravitational instability will typically dominate over the artificial viscosity, and so this choice of artificial viscosity is unlikely to strongly influence our results.} {To check this we also ran one of our more extreme models ($M_*=0.1\,$M$_\odot$, $M_d/M_*=1$, $R_d=200\,$AU, MIST irradiated) using $\alpha=0.1, \beta=2$  and another using the artificial viscosity reducing scheme of \cite{1997JCoPh.136...41M}. The result was always an axisymmetric disc, so the artificial viscosity scheme is not having any key influence upon our outcomes, as we expect from \cite{2011MNRAS.410..994F}.  }

\begin{figure}
    \centering
    \includegraphics[width=9cm]{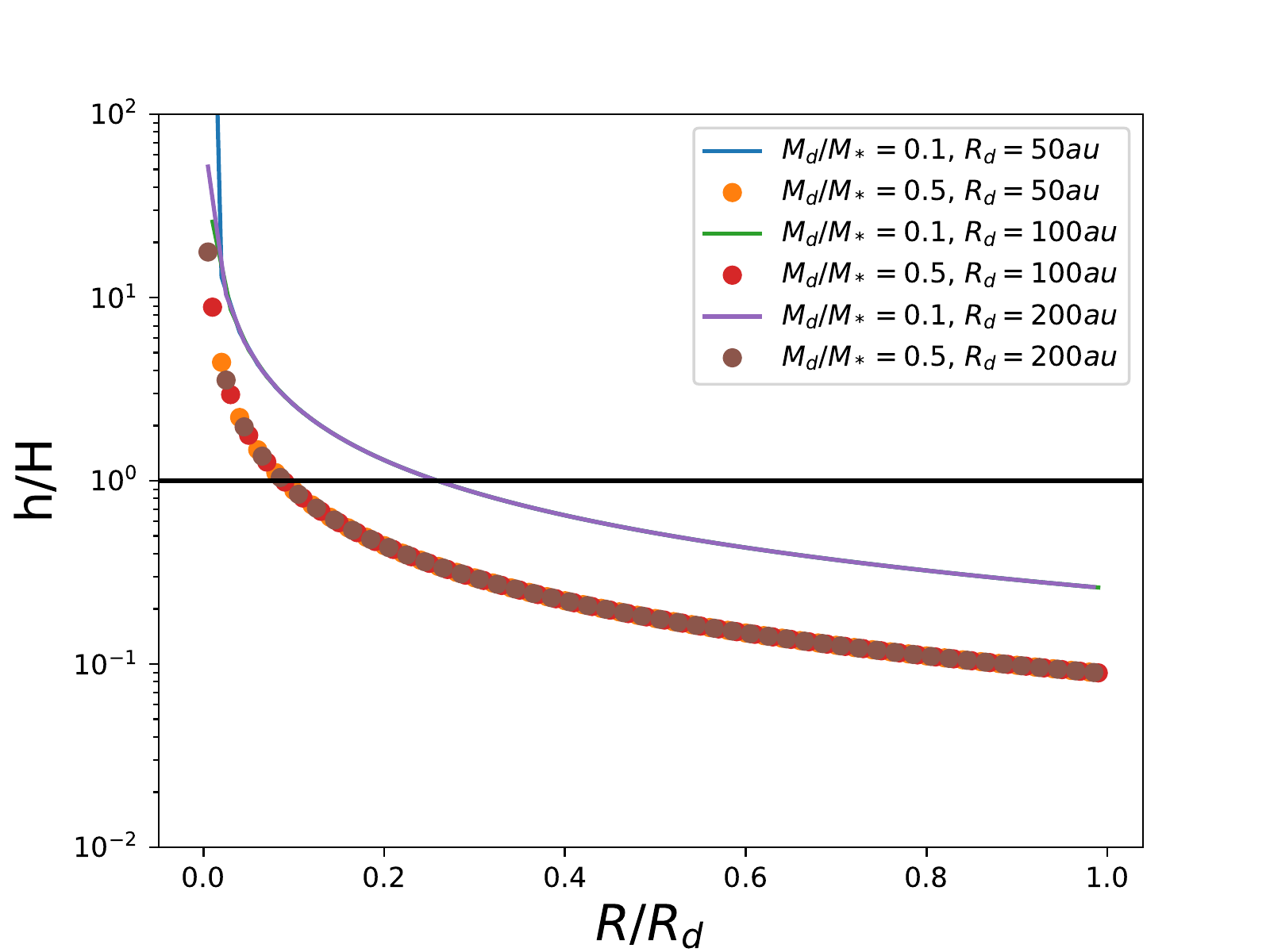}
    \caption{The ratio of particle smoothing length to disc scale height as a function of fraction of the total disc radius. {For our assumed initial surface density profile, $h/H$ is independent of the stellar mass and disc size, depending only on the disc-to-star mass ratio.} }
    \label{fig:resolution}
\end{figure}

\subsection{Resolution requirements}
\label{sec:resolution}
Our calculations are only required to discern between axisymmetric discs, those that settle into a quasi-steady state where spiral density waves act to transport angular momentum, and those that fragment. Since there are over 300 models in this paper (see section \ref{sec:params}) it is pragmatic to limit the resolution as much as possible without compromising the gross stability \citep[e.g.][]{2012MNRAS.427.2022M, 2015MNRAS.451.3987Y}. {We are not concerned with the detailed nature of the instability or disc substructure, which can be very sensitive to the resolution employed \citep{2012MNRAS.427.2022M}, just whether it deviates from axisymmetry, which is less so.}

From equation 14 of \cite{2012MNRAS.427.2022M} (and the associated discussion in that paper) the criterion for the disc being sufficiently resolved at radius $R$ is that the ratio of the SPH particle smoothing length $h$ to the scale height $H$
\begin{equation}
    \frac{h}{H}(R) \approx \frac{1.2}{\Sigma(R)R^2}\left(\frac{2M_*^2M_d}{\pi^2 N_{part}}\right)^{1/3}
    \label{equn:resolution}
\end{equation}
is $<\sim0.25$ \citep{ANelson2006} for stellar mass $M_*$ and disc mass $M_d$. {Note that for our assumed \textit{initial} surface density profile which scales as $R^{-1}$ this reduces to}
\begin{equation}
     \frac{h}{H}(R) \approx 2.4 \left(\frac{2\pi M_*^2}{M_d^2N_{part}}\right)^{1/3}
\end{equation}
{and hence has no dependency upon the disc size, only the disc-to-star mass ratio (of course the simulation will evolve away from these initial conditions). }

{We use $N_{part}=0.5$\,M particles for all of our calculations. In Figure \ref{fig:resolution} we plot $h/H$ for a selection of star/discs. For disc-to-star mass ratios of 0.1 and 0.5 the outer 75 and 90\,per cent of the disc has $h/H<1$ respectively. It is the outer regions where instability arises and so are the most important parts of the discs to resolve.  Because our calculations are, by necessity, not extremely well resolved, in Appendix \ref{sec:resConvTest} we run a convergence check for part of our grid to demonstrate that our usage of 0.5\,M particles is sufficient. We also demonstrate in the Appendix that the heating from stellar irradiation substantially reduces $h/H$ and those models are hence better resolved.}

\subsection{Model parameters}
\label{sec:params}
The discs in our models are defined in terms of their initial mass, size and minimum Q parameter. The latter sets the sound speed. {The disc inner radius is 1\,AU (we are not concerned with the very inner disc, since self-gravity influences the outer disc)}.  We initially impose a surface density of the form
\begin{equation}
   \Sigma=\Sigma_{1\textrm{AU}}\left(\frac{R}{\textrm{AU}}\right)^{-1}
\end{equation}
where $\Sigma_{1\textrm{AU}}$ is set by the disc mass and size
\begin{equation}
    \Sigma_{1\textrm{AU}} = \frac{M_d}{2\pi \left(R_d\times1\textrm{AU} - \textrm{AU}^2\,\right)}. 
\end{equation}
For reference we find that the resulting surface density profile in the 1D models of \ref{sec:timedep1D} goes roughly as $R^{-3/2}$, so it is similarly steep. {In a companion paper, Cadman et al. (submitted), this steeper surface density profile motivated by the 1D models is used and the conclusions drawn in this paper still hold. }
We assume an initial sound speed profile of the form 
\begin{equation}
   c_s=c_{s,0}\left(\frac{R}{\textrm{AU}}\right)^{-0.25}
\end{equation}
though of course in the calculations where the stellar passive irradiation is accounted for there can be a substantial deviation from this initial imposed temperature structure. The value of the initial minimum $Q$ value sets the initial thermal structure {(not accounting for the stellar irradiation)}. We set a minimum of $Q=2$ in our initial model setup. Of course in the irradiated cases the temperatures associated with this initial $Q$ choice will be rapidly overwritten.  A summary of parameters of our suite of simulations is given in Table \ref{tab:parameters}. 

We run each of our calculations either until they are clearly unstable, or for at least 3 Keplerian orbital timescales at the disc outer edge. {Once the initial grid was completed we also ran the most massive axisymmetric discs for a factor 2 longer (i.e. to at least 6 Keplerian orbital timescales at the disc outer edge) to ensure that substructure due to self-gravity was not missed due to a longer cooling timescale.}  {We checked the maximum thermal timescale of all particles for our most massive axisymmetric discs using equation 25 of \cite{2009MNRAS.394..882F}, finding that even in our most extreme model ($M_*=0.1$\,M$_\odot$, $M_d/M_*=1.4$, $R_d=200$\,AU) we ran for a factor 2 longer than the thermal timescale ($t_{\textrm{therm}}=$27\,kyr), and for most calculations the maximum thermal timescale is much shorter (<1kyr). }  In all of our models we use an inner (accretion) radius for the disc of 1\,AU.

{\cite{2011MNRAS.410..994F} found that global gravitational effects become important for $M_d/M_* > 0.5$, which was the upper limit in our initial grid of models. For calculations that were still stable we extended the grid to higher disc-to-star mass ratios which hence have the caveat that setting up disc-like initial conditions is probably unrealistic. However, given that these simulations are self-consistent they will still settle into states that are consistent with this large disc-to-star mass ratio.}

\begin{table*}
\caption{Summary of the grid of models used in this paper. Columns are the stellar mass, stellar luminosity from MIST, \protect\cite{2000A&A...358..593S} and whether we also consider a non-irradiated case. Next is the disc-to-star mass ratio, followed by the initial disc size, initial disc surface density  and the orbital period at the disc outer edge.   }
\label{tab:parameters}
\begin{tabular}{@{}lccccccccccc@{}}
 \hline
 $M_*$ & $L_*$ (MIST) & $L_*$ (Siess) &Non- &  $M_d/M_*$ &  $R_d$ &  $t_d$ &   \\
 $M_\odot$& $\log_{10}L_\odot$ & $\log_{10}L_\odot$  &irradiated case & & AU & kyr \\
  \hline
1.0 &0.52 &0.013&\checkmark  & 0.1 & 50, 100, 200    & (0.36, 1.0, 2.85)  \\
1.0 &0.52 &0.013&\checkmark  & 0.2 & 50, 100, 200   &  (0.36, 1.0, 2.85) \\
1.0 &0.52 &0.013&\checkmark  & 0.3 & 50, 100, 200  &  (0.36, 1.0, 2.85) \\
1.0 &0.52 &0.013&\checkmark  & 0.4 & 50, 100, 200  &  (0.36, 1.0, 2.85) \\
1.0 &0.52 &0.013&\checkmark  & 0.5 & 50, 100, 200  &  (0.36, 1.0, 2.85) \\
1.0&0.52 & --  & -- & 0.6&  200   &  2.85 \\
0.7 &0.28 &-0.019&\checkmark  & 0.1 & 50, 100, 200  &  (0.43, 1.21, 3.41)  \\
0.7 &0.28 &-0.019&\checkmark  & 0.2 & 50, 100, 200  & (0.43, 1.21, 3.41) \\
0.7 &0.28 &-0.019&\checkmark  & 0.3 & 50, 100, 200  &  (0.43, 1.21, 3.41) \\
0.7 &0.28 &-0.019&\checkmark  & 0.4 & 50, 100, 200  & (0.43, 1.21, 3.41) \\
0.7 &0.28 &-0.019&\checkmark  & 0.5 & 50, 100, 200  & (0.43, 1.21, 3.41) \\
0.7 &0.28 &--&--  & 0.6 & 200  &  3.41 \\
0.5 &0.058 &-1.27&\checkmark  & 0.1 & 50, 100, 200    &  (0.50, 1.43, 4.04)  \\
0.5 &0.058 &-1.27&\checkmark  & 0.2 & 50, 100, 200  &  (0.50, 1.43, 4.04) \\
0.5 &0.058 &-1.27&\checkmark  & 0.3 & 50, 100, 200  &  (0.50, 1.43, 4.04) \\
0.5 &0.058 &-1.27&\checkmark  & 0.4 & 50, 100, 200  &(0.50, 1.43, 4.04) \\
0.5 &0.058 &-1.27&\checkmark  & 0.5 & 50, 100, 200  & (0.50, 1.43, 4.04) \\
0.5 &0.058 &--&--  & 0.6 &  200  & 4.04 \\
0.3 &-0.24 &-1.77&\checkmark  & 0.1 & 50, 100, 200  &   (0.65, 1.84, 5.21)  \\
0.3 &-0.24 &-1.77&\checkmark  & 0.2 & 50, 100, 200  &  (0.65, 1.84, 5.21) \\
0.3 &-0.24 &-1.77&\checkmark  & 0.3 & 50, 100, 200  &  (0.65, 1.84, 5.21) \\
0.3 &-0.24 &-1.77&\checkmark  & 0.4 & 50, 100, 200  &  (0.65, 1.84, 5.21) \\
0.3 &-0.24 &-1.77&\checkmark  & 0.5 & 50, 100, 200  &    (0.65, 1.84, 5.21) \\
0.3 &-0.24 &--&--  & 0.6 &  100, 200  &  (1.84, 5.21) \\
0.3 &-0.24 &--&--  & 0.7 &  200  &5.21 \\
0.3 &-0.24 &--&--  & 0.8 &  200  &  5.21 \\
0.2 &-0.56 &-2.13&\checkmark  & 0.1 & 50, 100, 200  &   (0.80, 2.26, 6.38)  \\
0.2 &-0.56 &-2.13&\checkmark  & 0.2 & 50, 100, 200  &  (0.80, 2.26, 6.38)  \\
0.2 &-0.56 &-2.13&\checkmark  & 0.3 & 50, 100, 200  &   (0.80, 2.26, 6.38)  \\
0.2 &-0.56 &-2.13&\checkmark  & 0.4 & 50, 100, 200  &   (0.80, 2.26, 6.38)  \\
0.2 &-0.56 &-2.13&\checkmark  & 0.5 & 50, 100, 200  &   (0.80, 2.26, 6.38)  \\
0.2 &-0.56 &-2.13&--  & 0.6 & 50, 100, 200  & (0.80, 2.26, 6.38)  \\
0.2 &-0.56 &--&--  & 0.7 & 100, 200  &   (2.26, 6.38)  \\
0.2 &-0.56 &--&--  & 0.8 &  200  &   6.38 \\
0.2 &-0.56 &--&--  & 0.9 & 200  &   6.38 \\
0.2 &-0.56 &--&-- & 1.0 & 200  &   6.38 \\
0.1 &-1.13 &-3.08&\checkmark  & 0.1 & 50, 100, 200  &  (1.13, 3.19, 9.03)  \\
0.1 &-1.13 &-3.08&\checkmark  & 0.2 & 50, 100, 200  &   (1.13, 3.19, 9.03) \\
0.1 &-1.13 &-3.08&\checkmark  & 0.3 & 50, 100, 200  &   (1.13, 3.19, 9.03) \\
0.1 &-1.13 &-3.08&\checkmark  & 0.4 & 50, 100, 200  &  (1.13, 3.19, 9.03) \\
0.1 &-1.13 &-3.08&\checkmark  & 0.5 & 50, 100, 200  &   (1.13, 3.19, 9.03) \\
0.1 &-1.13 &-3.08&\checkmark  & 0.6 & 50, 100, 200  &  (1.13, 3.19, 9.03) \\
0.1 &-1.13 &-3.08&\checkmark  & 0.7 & 50, 100, 200  &   (1.13, 3.19, 9.03) \\
0.1 &-1.13 &-3.08&--  & 0.8 & 50, 100, 200  &   (1.13, 3.19, 9.03) \\
0.1 &-1.13 &-3.08&-- & 0.9 &  100, 200   &  (3.19, 9.03) \\
0.1 &-1.13 &-3.08&--  & 1.0 &  100, 200   &  (3.19, 9.03) \\
0.1 &-1.13 &--&--  & 1.1 &  100, 200  &   (3.19, 9.03) \\
0.1 &-1.13 &--& -- & 1.2 &  200  &  9.03 \\
0.1 &-1.13 &--&--  & 1.3 &  200  &  9.03 \\
0.1 &-1.13 &--&--  & 1.4 &  200  &  9.03 \\

 \hline
 \hline
\end{tabular}
\end{table*}

\section{Results}

\begin{figure*}
    \centering
    \includegraphics[height=0.6cm]{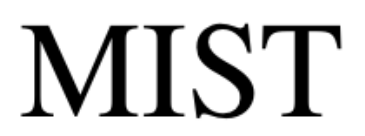}
    \vspace{-0.05cm}
    
    \hspace{-0.8cm}
    \includegraphics[width=6.2cm]{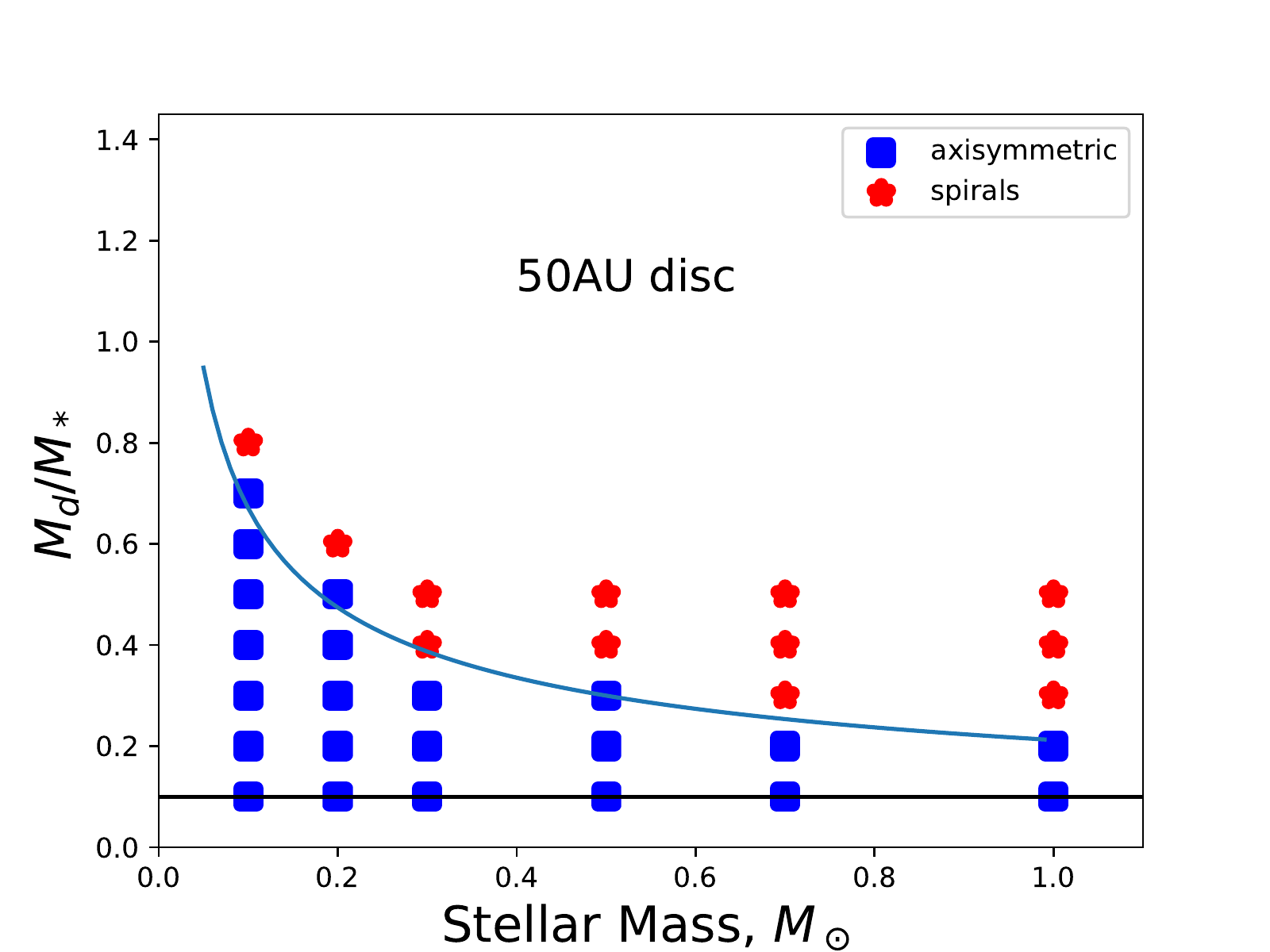}
    \hspace{-0.5cm}    
    \includegraphics[width=6.2cm]{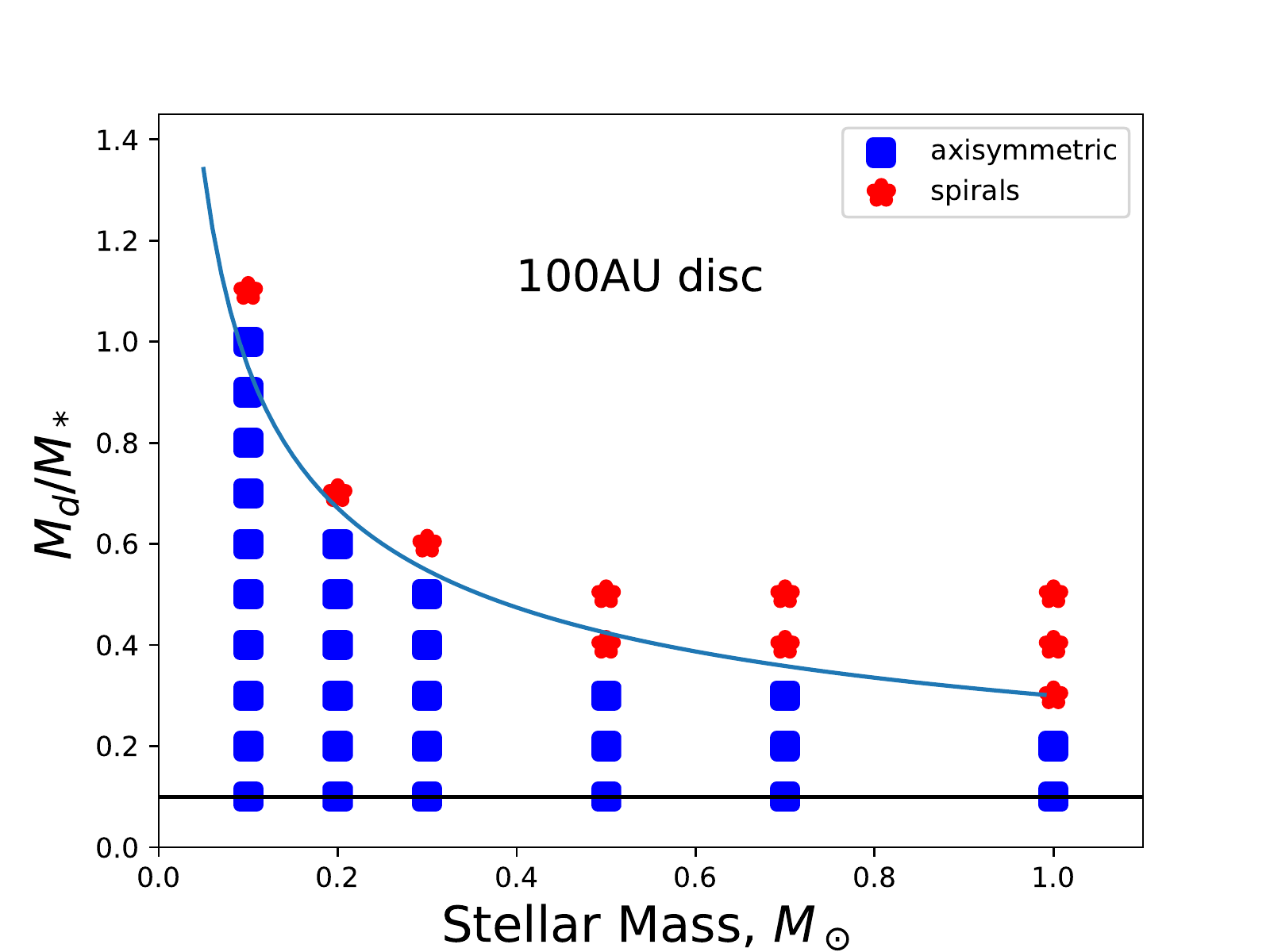}    
    \hspace{-0.5cm}   
    \includegraphics[width=6.2cm]{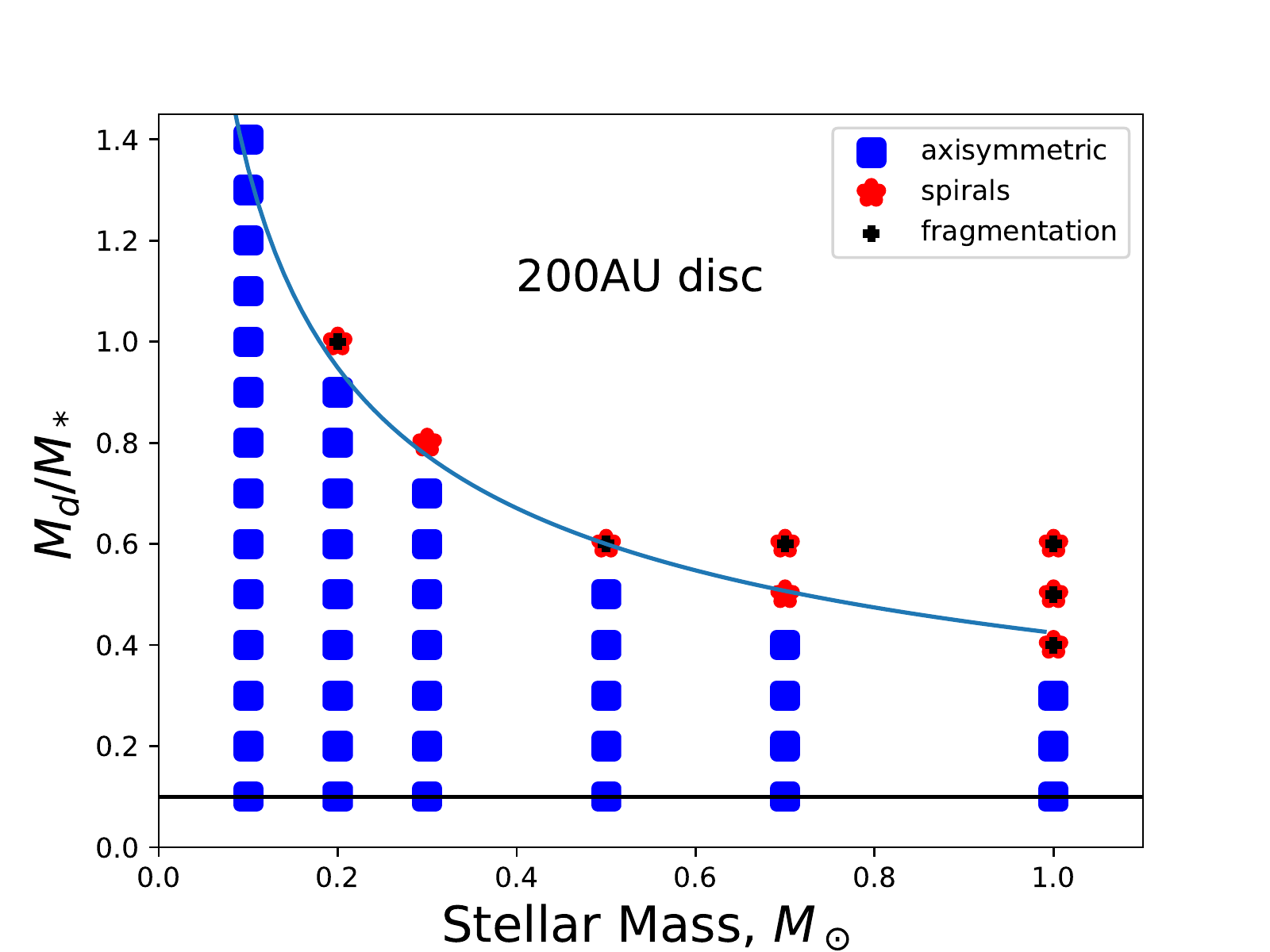}     

    \vspace{0.5cm}
      \includegraphics[height=0.6cm]{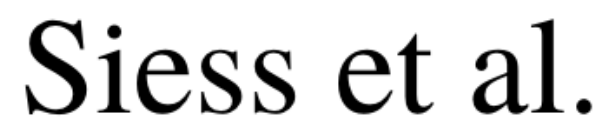}
    \vspace{-0.05cm}
    
    \hspace{-0.8cm}
    \includegraphics[width=6.2cm]{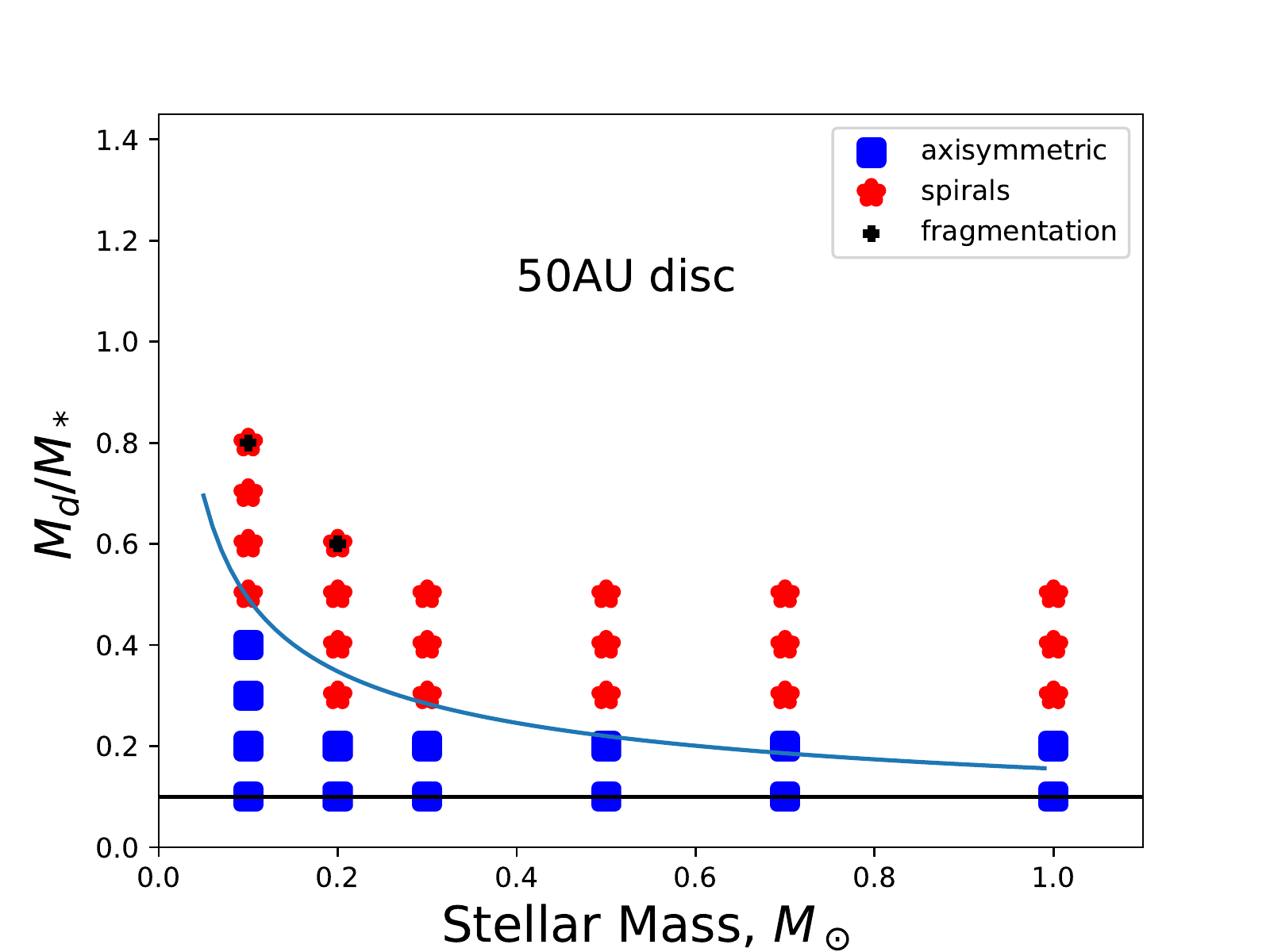}
    \hspace{-0.5cm}    
    \includegraphics[width=6.2cm]{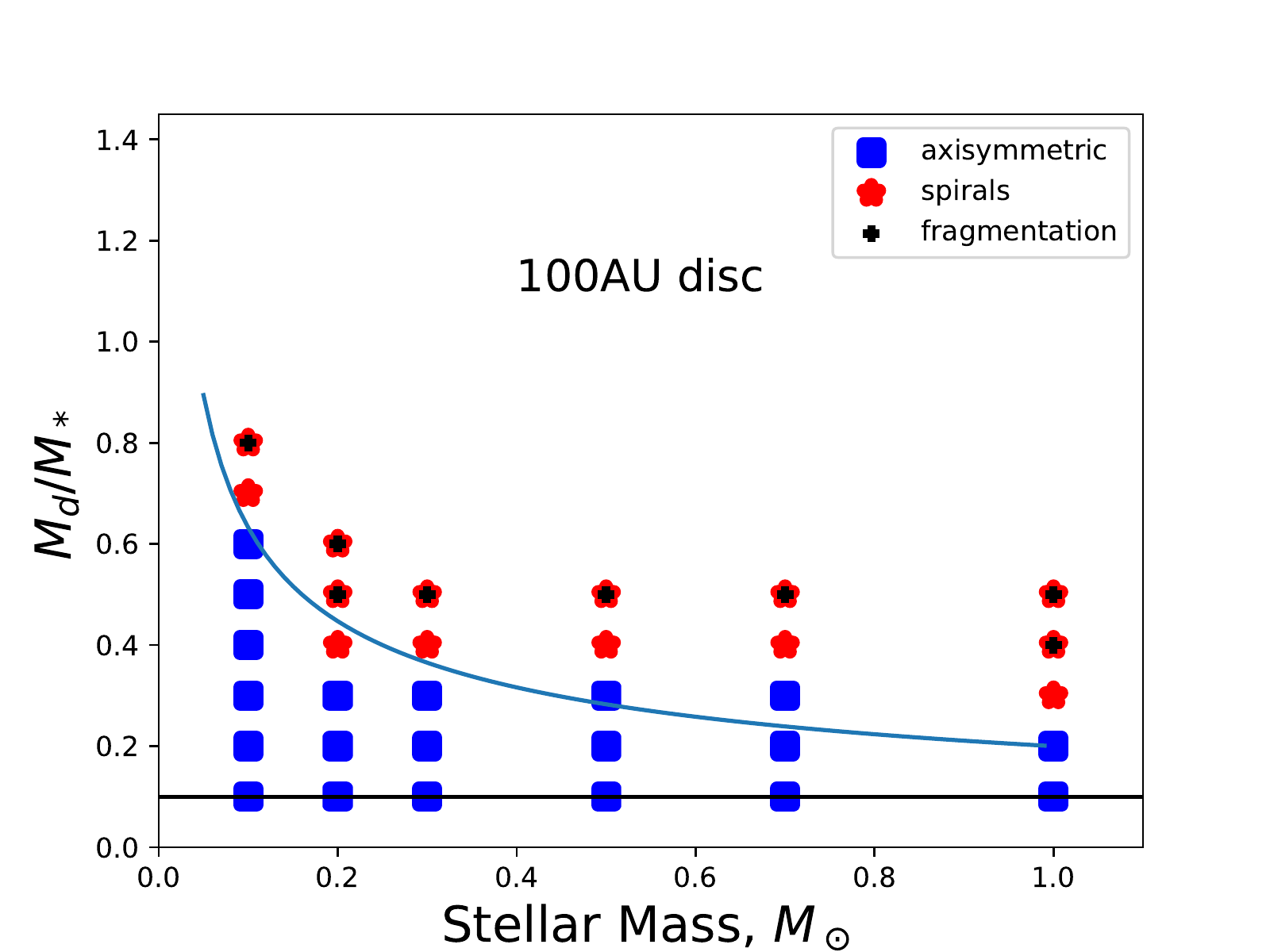}    
    \hspace{-0.5cm}   
    \includegraphics[width=6.2cm]{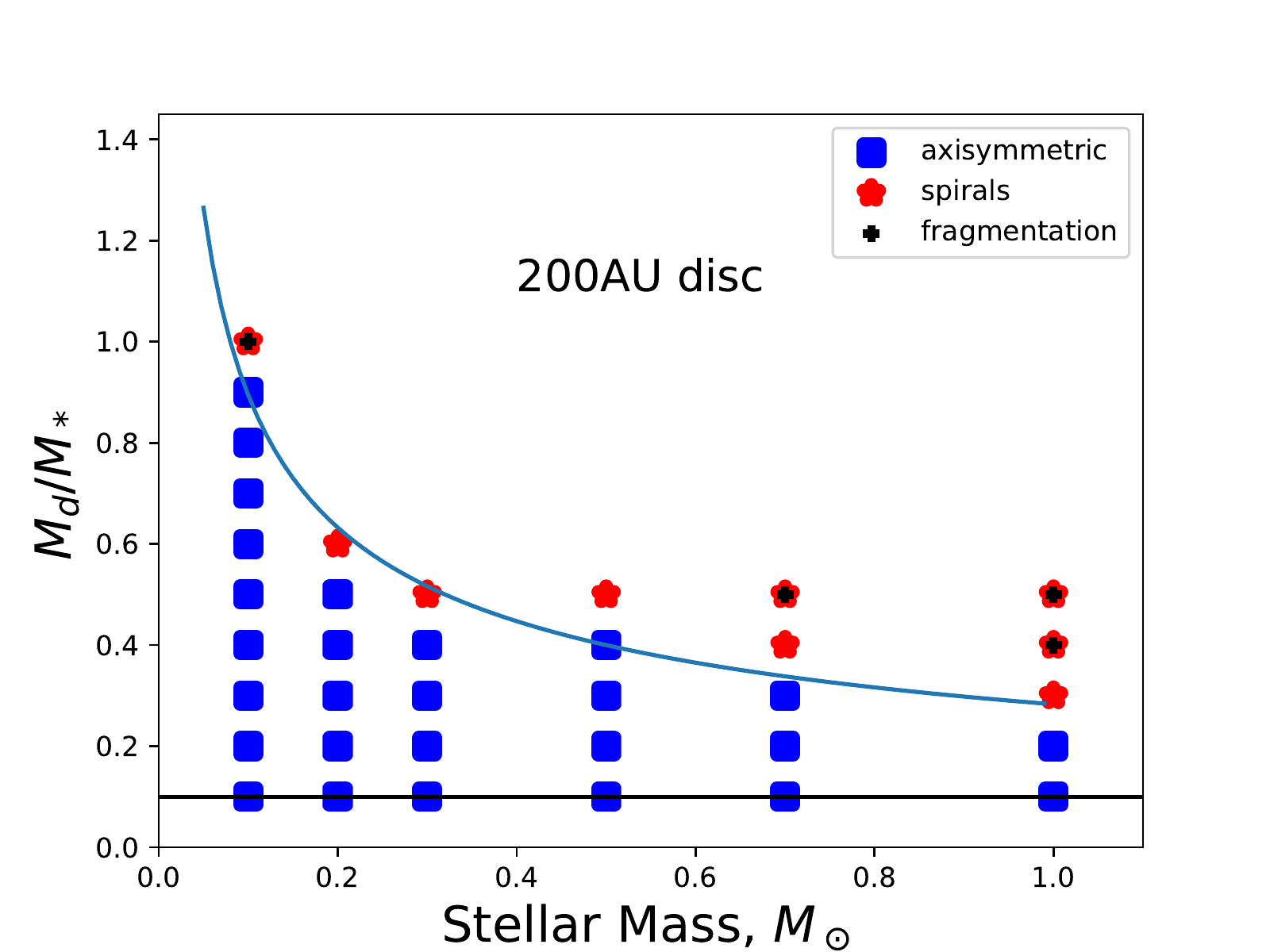}

    \vspace{0.5cm}
        \includegraphics[height=0.6cm]{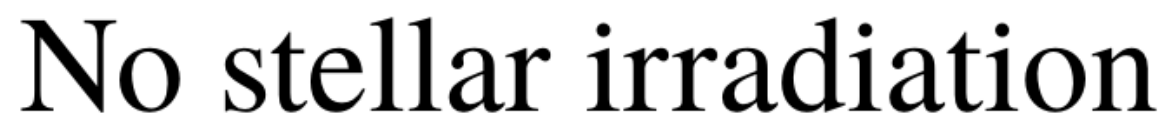}
        \vspace{-0.05cm}
        
        \hspace{-0.8cm}
    \includegraphics[width=6.2cm]{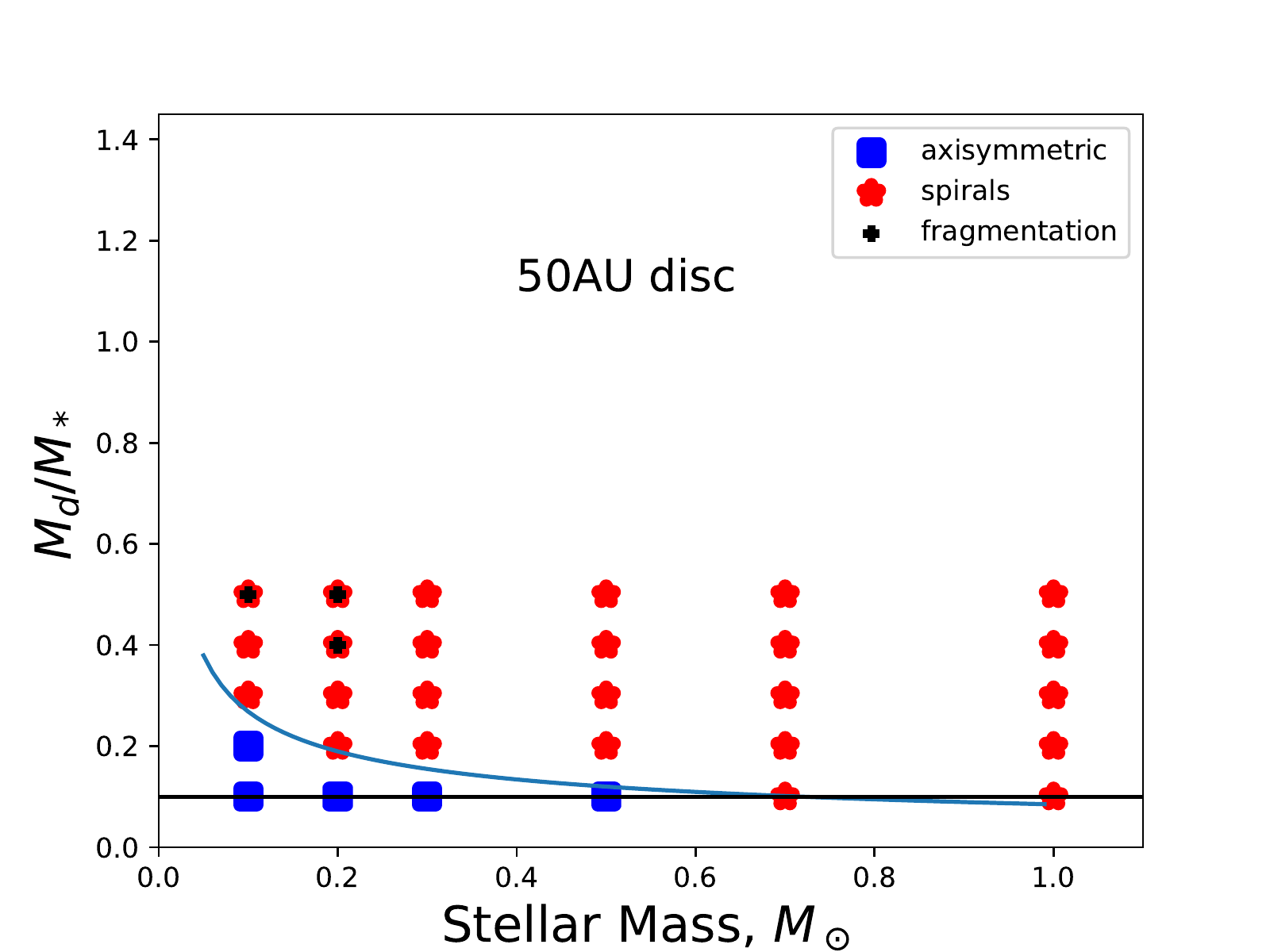}
    \hspace{-0.5cm}    
    \includegraphics[width=6.2cm]{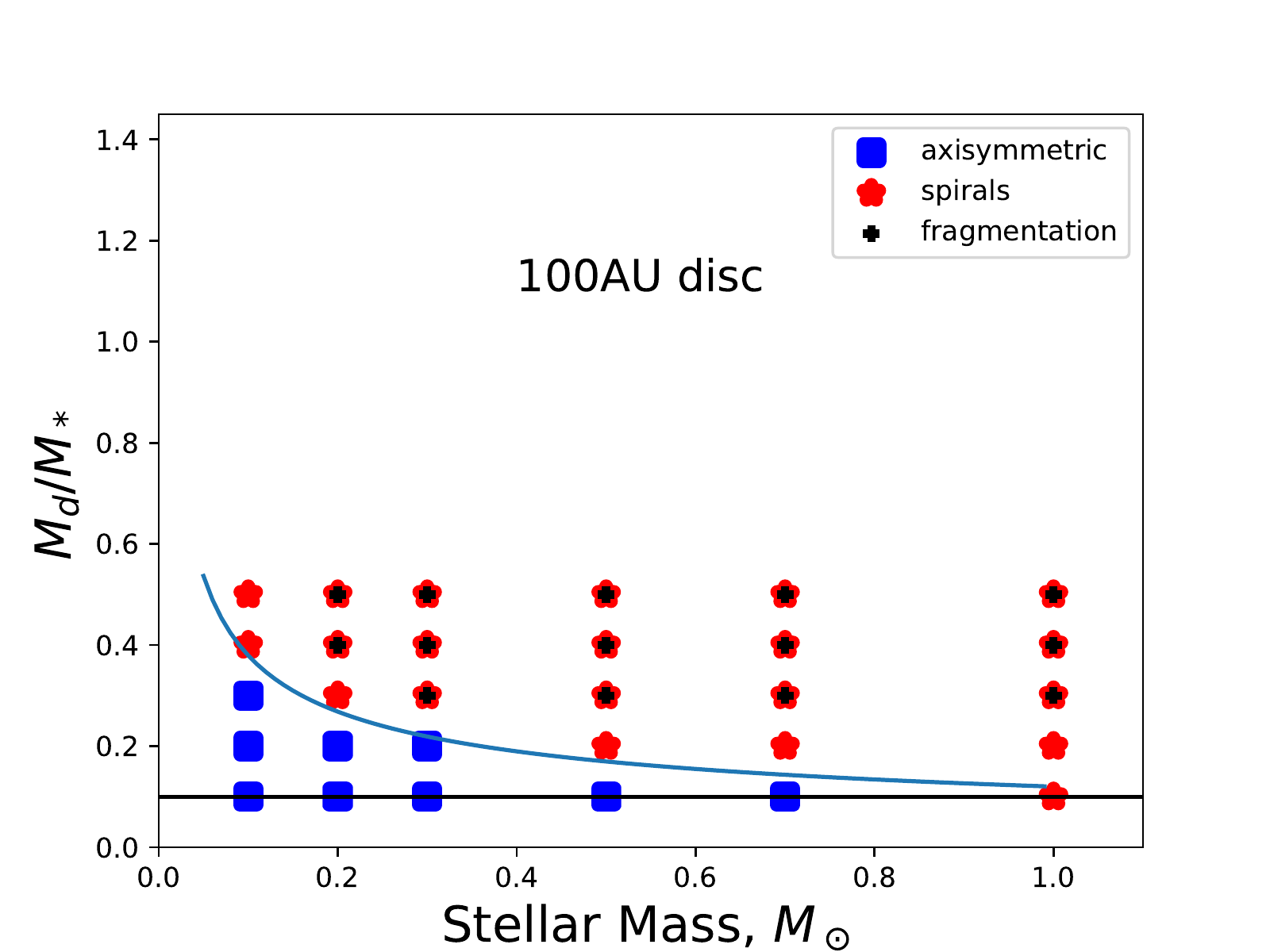}    
    \hspace{-0.5cm}   
    \includegraphics[width=6.2cm]{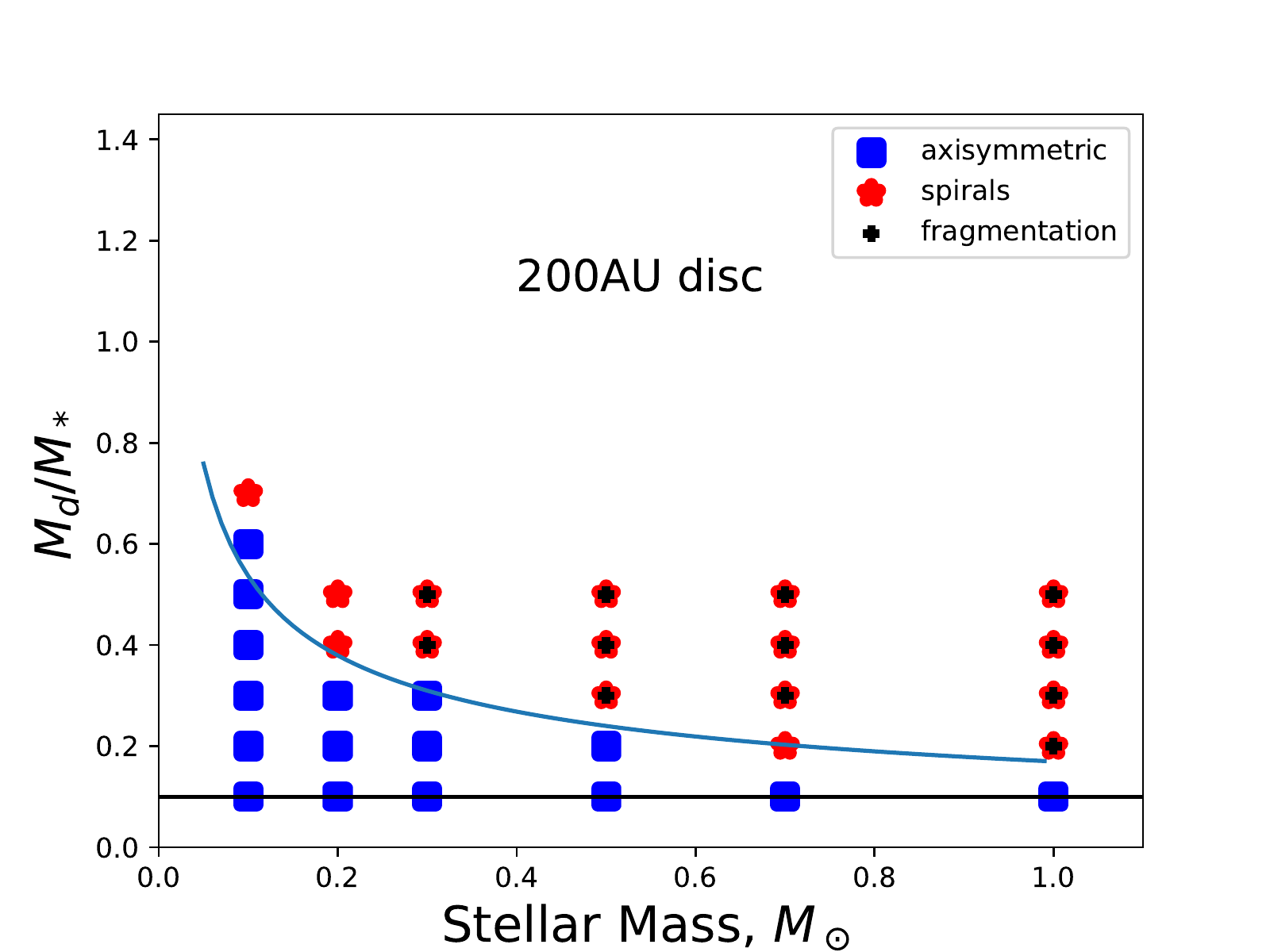}

    \caption{The primary summary of our model grid. Each panel shows the disc-to-star mass ratio and stellar mass of our models. The blue (square) points are models with stable discs, red (stars) are unstable. Those that also include a black point have undergone fragmentation. The upper row use 0.5\,Myr MIST models for the stellar irradiation, the second use \protect\cite{2000A&A...358..593S} ZAMS models and the lower panels do not include passive irradiation by the host star. The curves which approximately distinguish between axisymmetric and spiral models are given by equations  \ref{equn:irr_ratioMIST}--\ref{equn:irr_rationoIrr}.}
    \label{fig:mainSummary}
\end{figure*}

\subsection{Overview of SPH model grid}
\label{sec:overview}
We begin with a general overview of our grid of models. The main summary of the behaviour of discs in our grid is given in Figure \ref{fig:mainSummary}. Each panel consists of points representing the disc-to-star mass ratio and stellar mass of each model. Blue squares are axisymmetric discs, red stars are models that exhibit spiral features (i.e. they are marginally stable with $Q\sim1.7)$. The models that are sufficiently unstable for fragmentation to occur are further marked with a black cross. Note that we manually characterize our models, since {the difference between spirals and fragmentation} is unsubtle. Also recall that our stable models are run for at least a few orbital timescales of the disc outer edge. In Figure \ref{fig:mainSummary}, the blue line represents a criterion for separating axisymmetric/spiral discs that we will quantify in section \ref{sec:scaling}.

The upper panels of Figure \ref{fig:mainSummary} use the 0.5\,Myr MIST stellar luminosities, the middle panels the ZAMS \cite{2000A&A...358..593S} stellar luminosities and the lower panels do not include passive stellar irradiation (refer back to Figure \ref{fig:luminosities} and Table \ref{tab:parameters} for a summary of the stellar mass-luminosity functions we use, but note that the passive irradiation decreases from top to botton in Figure \ref{fig:mainSummary}). 

In all cases the behaviour is consistent with  \cite{2016ARA&A..54..271K} (equation \ref{equn:KratterLodato} in this paper) and the expectation from our 1D time dependent models discussed in Sections \ref{sec:timedep1D}, and \ref{sec:1Dacc}. We will quantify the criteria for axisymmetry further in section \ref{sec:scaling}, but broadly as the stellar mass decreases, the maximum stable disc-to-star mass ratio increases. At the lowest stellar masses (approaching that of Trappist-1) this translates into significantly higher axisymmetric disc-to-star mass ratios of 0.5 or even higher. Furthermore larger discs are also capable of supporting higher mass axisymmetric discs. The final effect is that there is increasing stability with the amount of passive heating (stellar luminosity), which we will next discuss in more detail. 

\begin{figure*}
    \centering
    \includegraphics[width=18cm]{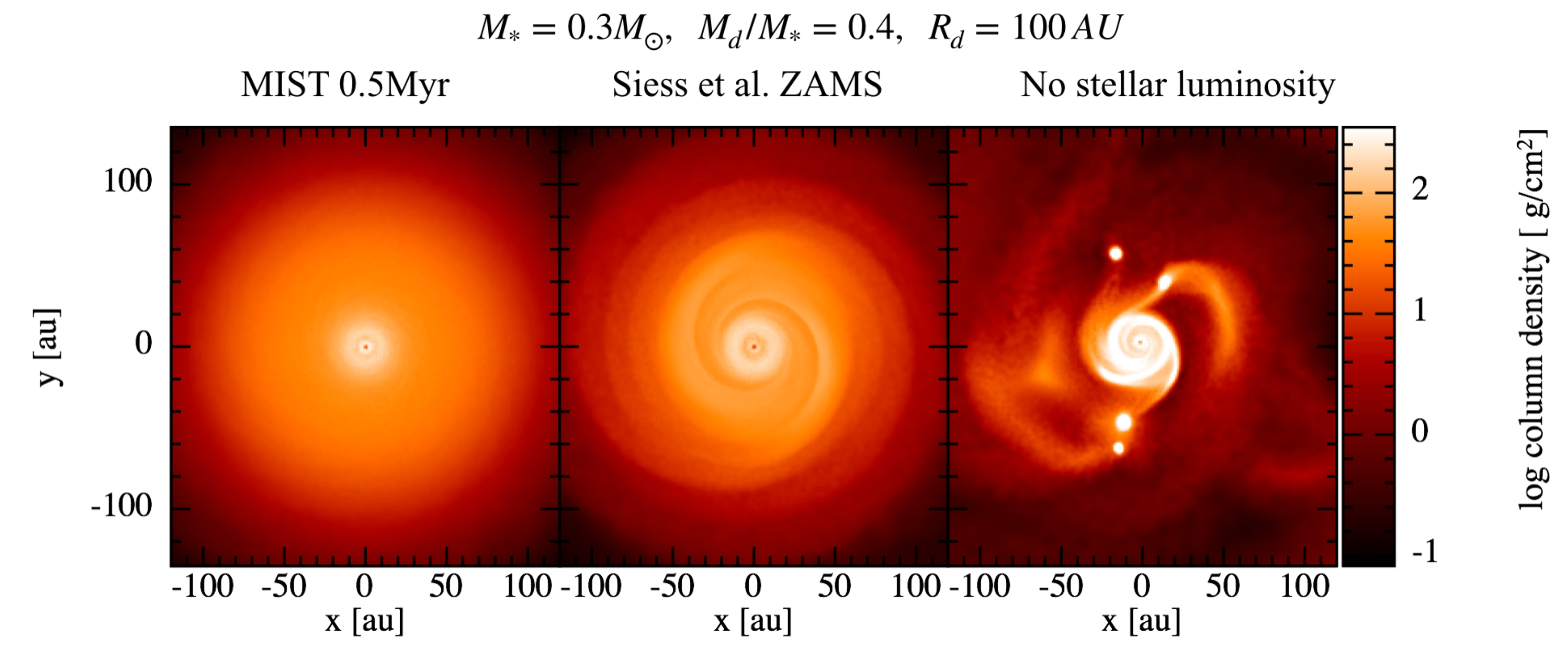}
    \caption{A comparison of disc evolution with different treatments of the stellar luminosity for a 0.3\,M$_\odot$ star with a disc-to-star mass ratio $M_dM_*=0.4$ and a disc radius of 100\,AU. The left hand panel includes stellar heating using a luminosity from the MIST evolutionary tracks at 0.5\,Myr. The central panel uses the (comparatively lower) ZAMS luminosities from \protect\cite{2000A&A...358..593S} and the right hand panel includes no stellar irradiation. All that differs is the treatment of stellar irradiation, but the impact that this has upon the dynamical evolution can differ dramatically.  }
    \label{fig:varyLum}
\end{figure*}

\subsubsection{The impact of irradiation}
As discussed by, for example, \cite{2008ApJ...673.1138C}, \cite{2010MNRAS.406.2279M}, \cite{2011ApJ...740....1K}, \cite{2012MNRAS.423.1896R} \cite{2011MNRAS.418.1356R} and \cite{2016MNRAS.458..306H} irradiation  of a disc (by the host star or an external field) can play a key role in weakening or completely suppressing the gravitational instability in discs and this is reflected by our grid. 

Our models without any stellar heating are pervasively exhibiting spirals and/or unstable {to fragmentation}. The additional heating from passive stellar irradiation in the ZAMS \cite{2000A&A...358..593S} models suppresses spirals and fragmentation by increasing the thermal support against collapse (equivalently,  the sound speed increases in the Toomre Q  stability parameter, equation \ref{equn:simpleQ}). This effect is even stronger in the 0.5\,Myr MIST stellar models, which have higher luminosities that are also a relatively flat function of  stellar mass (see Figure \ref{fig:luminosities}). Referring to Figure \ref{fig:mainSummary}, passive stellar irradiation plays an important role in reducing gravitational instability even for Solar like stars, where the difference between no irradiation and the 0.5\,Myr MIST stellar luminosity is axisymmetry for  disc-to-star mass ratio of $\leq0.1$ and $\leq0.3$ respectively. 

\begin{figure}
    \centering
    \includegraphics[width=8.7cm]{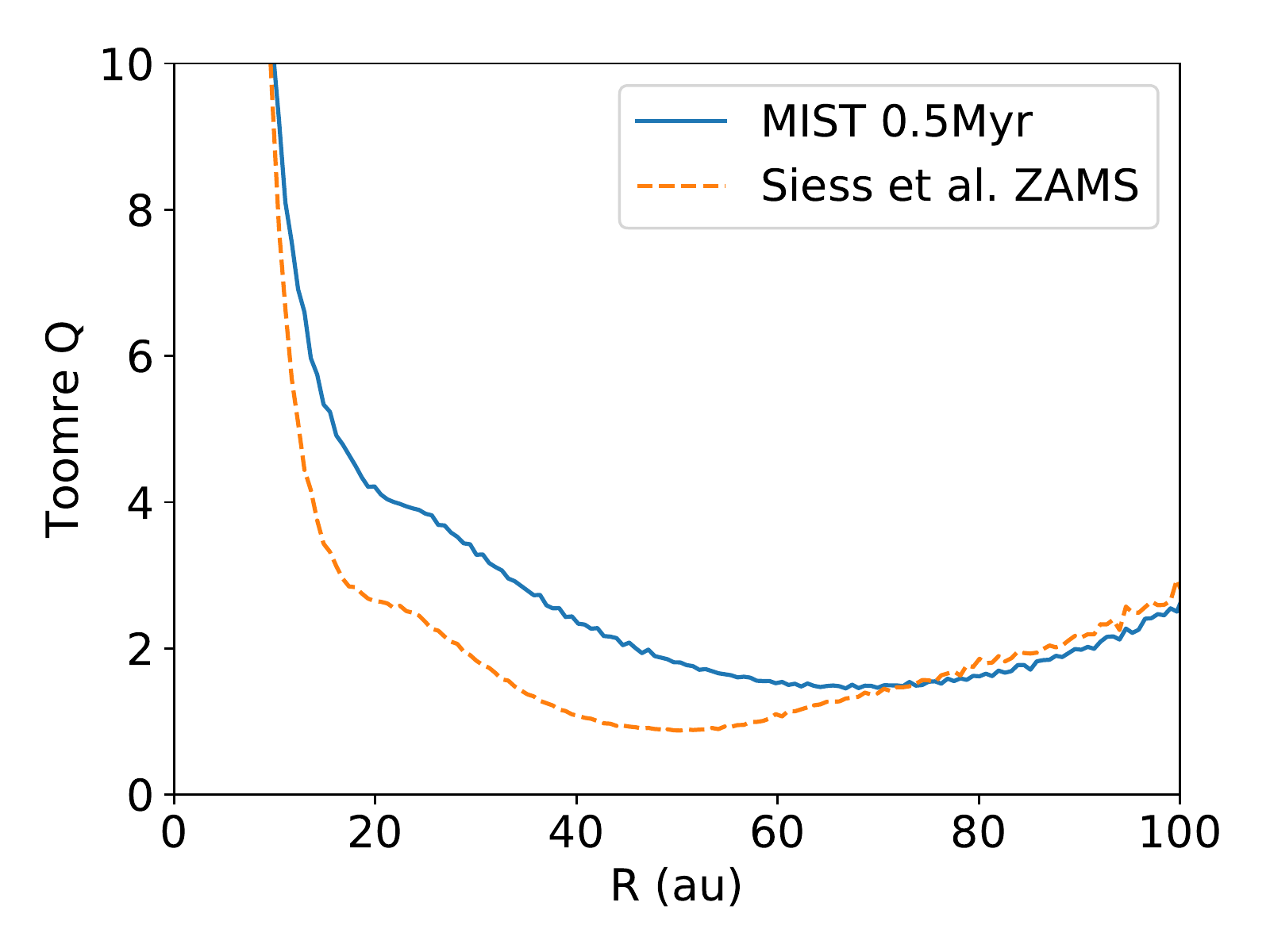}
    \includegraphics[width=8.7cm]{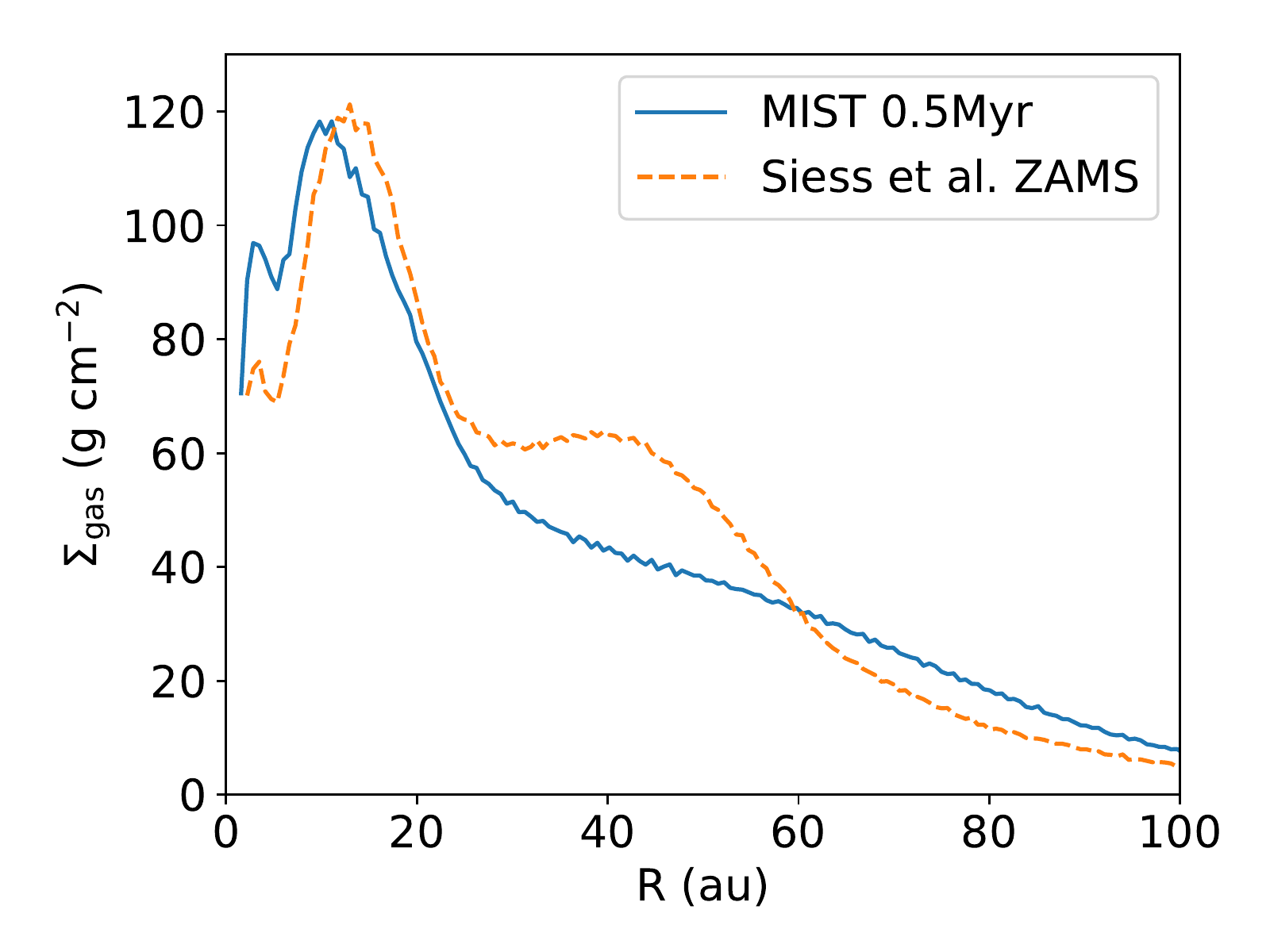}  
    \includegraphics[width=8.7cm]{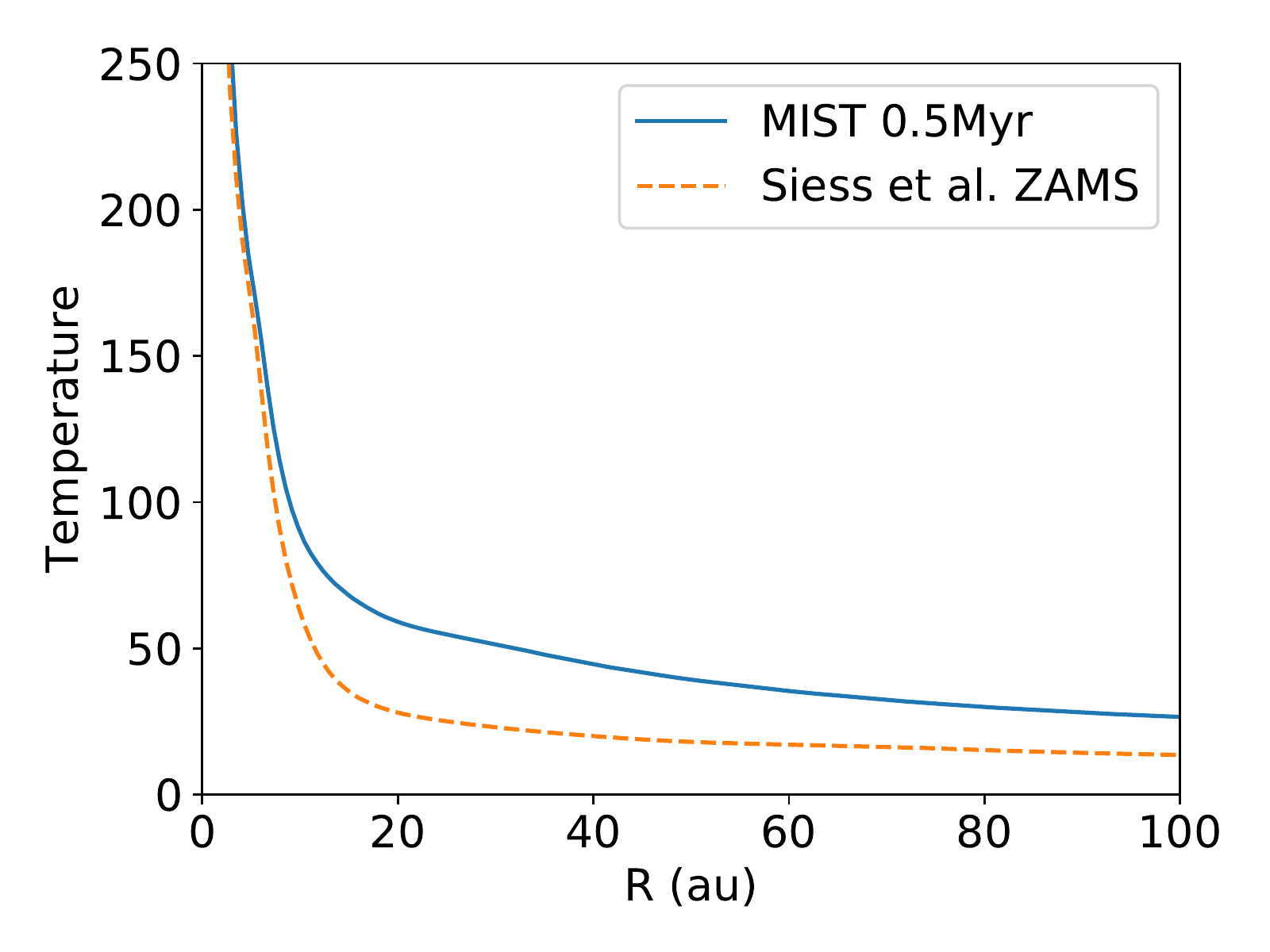}   
    \vspace{-0.2cm}
    \caption{{The upper panel is the azimuthally averaged Toomre Q parameter profiles for the left and central panels of Figure \ref{fig:varyLum} (blue solid and orange dashed lines of this Figure respectively). The blue solid has $Q>1.5$ everywhere and is axisymmetric. The orange dashed has $Q\sim1$ and exhibits quasi-steady spirals. The central and lower panels are the azimuthally averaged  surface density and temperature profile for these models. }}
    \label{fig:aziToomreQ}
\end{figure}

We further illustrate the potential importance of stellar irradiation in Figure \ref{fig:varyLum}, which shows the end state of disc models in the case of a 0.3\,M$_\odot$ star and a disc-to-star mass ratio of $M_d/M_* \sim 0.4$. All that differs between each panel is the stellar irradiation, with the left panel using the 0.5\,Myr MIST luminosities, the central panel the \cite{2000A&A...358..593S} luminosities and the right hand panel includes no stellar irradiation. For these particular star-disc parameters the difference in stellar irradiation alone is enough to change the evolutionary outcome from an axisymmetric disc, to quasi-steady spirals, to fragmentation, as the stellar heating decreases. Although it is clear that stellar irradiation plays a key role and the exact luminosity can also mean the difference between spirals and fragments, more generally across our grid the disc stability is similar for the \cite{2000A&A...358..593S} and MIST models. {For reference, we include the azimuthally averaged Toomre Q parameter {and surface density and temperature profile} for the MIST and ZAMS irradiated models of Figure \ref{fig:varyLum} in Figure \ref{fig:aziToomreQ}.}

\begin{figure*}
    \centering
    \includegraphics[width=18cm]{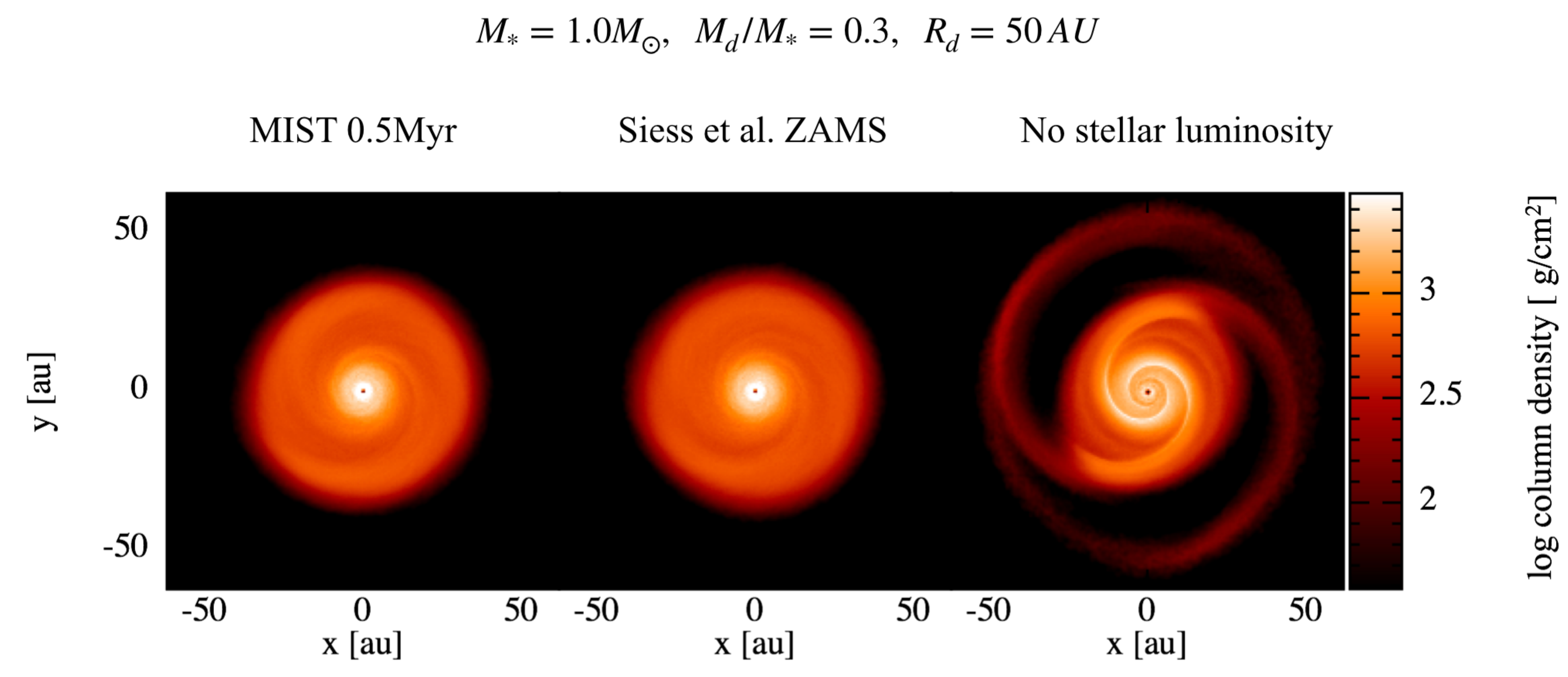}
    \caption{Spiral waves in three of our models. Each calculation has a Solar mass star and a 50\,AU 0.3\,M$_\odot$  disc. The right hand panel has no stellar passive heating and the left and central panels consider ZAMS and 0.5\,Myr stellar irradiation respectively. The enhanced heating reduces the amplitude of the spiral structure relative to the non-irradiated case.}
    \label{fig:varyLumSpirals}
\end{figure*}

In some cases on our grid the disc is classified as having spirals for all treatments of the stellar irradiation (or lack thereof), as illustrated in Figure \ref{fig:varyLumSpirals}. This shows models with MIST irradiation (left), \cite{2000A&A...358..593S} ZAMS irradiation (centre) and no irradiation (right) with the same star-disc parameters. Although all three are deemed ``unstable'' with spirals, the model without stellar irradiation has much higher amplitude spirals since there is less heating to oppose collapse.  So in addition to irradiation altering whether a disc is axisymmetric or not, it also influences the {amplitude} of the instability itself. 

Ultimately, stellar irradiation plays an important role, which is generally to reduce the gravitational instability. Although it seems that even a small amount of heating can provide this suppression \citep[as in the cases that use the][models]{2000A&A...358..593S}, our models show that it is actually possible that the magnitude of the stellar luminosity can make the difference between stability, spirals and fragmentation. 

We end by reiterating that our models do not account for self-shielding, or a possible large column {density (and hence opacity)} in the inner disc. These would potentially lower the irradiation heating and lessen its impact in stabilising discs. However the limiting case of this effect would just be our non-irradiated disc models.

\subsection{Scaling of critical disc-to-star mass ratio for axisymmetry}
\label{sec:scaling}
In section \ref{sec:overview} we discussed the broad properties of our grid, and have shown that the general behaviour is consistent with the expectations from our simpler one-dimensional analyses and from equation \ref{equn:KratterLodato}, taken from \cite{2016ARA&A..54..271K}. However the thermal term in equation \ref{equn:KratterLodato} is not trivially evaluated, making it of limited utility for evaluating the possible disc-to-star mass ratio of real systems that do/do not exhibit deviations from axisymmetry, like spirals and/or fragmentation. We provide this utility by providing axisymmetry criteria based on our model grids, for each treatment of the stellar luminosity.

For our primary models that consider 0.5\,Myr stars with MIST luminosities, the criterion on the disc-to-star mass ratio for an axisymmetric disc is reasonably described by 
\begin{equation}
    \frac{M_d}{M_*} < 0.3\left(\frac{R_d}{100\,\textrm{AU}}\right)^{1/2}\left(\frac{M_*}{M_\odot}\right)^{-1/2}
    \label{equn:irr_ratioMIST}
\end{equation}
where $R_d$ is the disc outer radius. This function is overplotted as the blue curve on the upper panels of Figure \ref{fig:mainSummary}. At very low stellar masses the agreement is less good, with discs actually more likely to be axisymmetric than this expression predicts. This further steepening of the function at very low stellar masses is due to the flatter mass--luminosity distribution at such an early stage in the stellar lifetime (i.e. there is higher luminosity at low stellar  mass in the models using the MIST luminosities compared to those on the ZAMS). However the impact upon the accuracy is relatively small and equation \ref{equn:irr_ratioMIST} still places much tighter constraints on the possible disc parameters than those that are usually applied, whilst remaining simple. Adding in a mass-dependent luminosity term in equation \ref{equn:irr_ratioMIST} to improve the agreement would add substantial complexity, reducing the utility, without significantly changing the quality of the criterion. We therefore favour equation \ref{equn:irr_ratioMIST}.

We did also automatically fit for the best parameters for these criteria and find similar values. However given the relatively sparse nature of the grid and uncertainties, e.g. in the inner optical depth, we choose to retain our manually constructed criteria.

Similarly, for our \cite{2000A&A...358..593S} ZAMS stellar luminosity models we find {that} a reasonable criterion for axisymmetry is given by
\begin{equation}
    \frac{M_d}{M_*} < 0.2\left(\frac{R_d}{100\,\textrm{AU}}\right)^{1/2}\left(\frac{M_*}{M_\odot}\right)^{-1/2}
    \label{equn:irr_ratioSiess}
\end{equation}
though we reiterate that the ZAMS luminosities are not expected at such an early time in the disc life and include this only to demonstrate how an intermediate luminosity distribution (e.g. due to an optically thick disc) alters the normalization. 
Lastly, for our non-irradiated discs. 
\begin{equation}
    \frac{M_d}{M_*} < 0.12\left(\frac{R_d}{100\,\textrm{AU}}\right)^{1/2}\left(\frac{M_*}{M_\odot}\right)^{-1/2}. 
    \label{equn:irr_rationoIrr}
\end{equation}

So overall the only real distinction between our criteria for the three different treatments of the stellar luminosity is the normalisation of the critical disc-to-star mass ratio function. The same functional form provides a reasonable description of the model behaviour in spite of the steeper mass-luminosity profile at very early times (refer to Figure \ref{fig:luminosities}). Equations  \ref{equn:irr_ratioMIST}--\ref{equn:irr_rationoIrr} are viable tools to place constraints on the modelling and interpretation of observations of real systems. 

\begin{figure}
    \includegraphics[width=8cm]{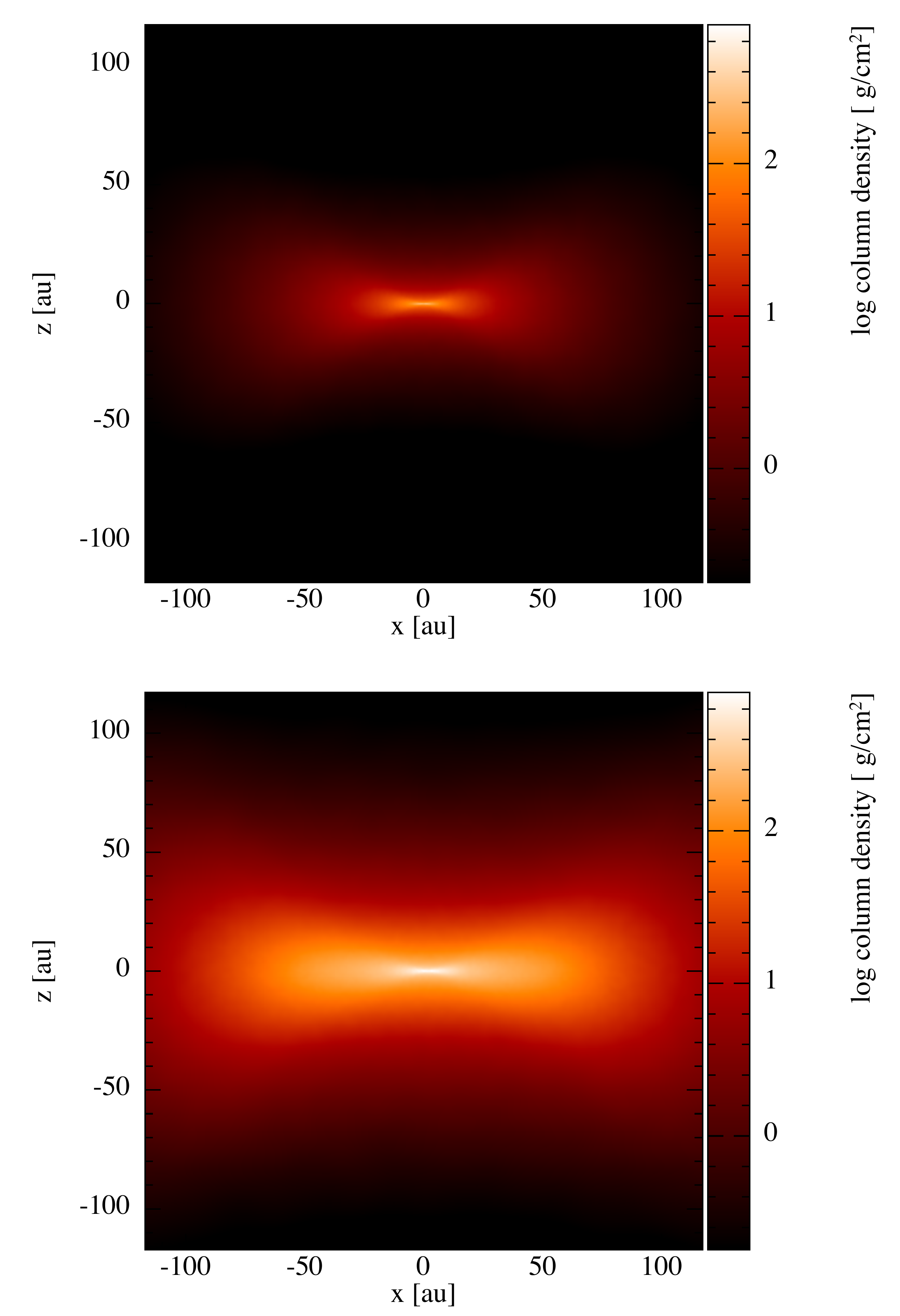}
    \caption{{A comparison of the edge-on view of the final steady state for 0.1\,$M_\odot$ star models, including MIST stellar irradiation, with 100AU discs of disc-to-star mass ratio 0.1 (top panel) and 1 (lower panel). Both are steady and axisymmetric, though the model in the lower panel is not formally a disc, with global gravitational effects becoming important as well as pressure support.} }
    \label{fig:EdgeOn}
\end{figure}

{We have already mentioned that for disc-to-star mass ratios above 0.5 that global gravitational effects can become significant \citep{2011MNRAS.410..994F}. In Figure \ref{fig:EdgeOn} we compare the edge-on view of the 0.1\,$M_\odot$ star model with MIST irradiation, a 100\,AU disc and disc-to-star mass ratios of 0.1 and 1 (both of which are axisymmetric). They share the same thermal scale height (since the temperature is dominated by stellar irradiation) and even significantly higher densities in the more massive disc do not lead to substantially different morphology}. {We note that in reality the extremely  high disc-to-star mass ratios (e.g. $>0.5$) will be difficult to achieve because it is unlikely that the star would begin so perfectly at at the centre of mass of the system as it does in these numerical experiments. This would lead to more complex dynamical evolution.  }

\subsection{How effective is stellar heating of the disc? }
\label{sec:radeq}
{Our criteria in equations \ref{equn:irr_ratioMIST} and \ref{equn:irr_rationoIrr} are based on the limiting cases in which the disc is optically thin and thick to the host star irradiation respectively. Since they vary in their normalization by a factor {of a} few, it will be important to know which extreme is most applicable in reality (in the absence of more detailed models that solve the radiative equilibrium temperature). Although addressing this fully is beyond the scope of this work, to make an initial assessment we constructed parametric static discs following \cite{2014ApJ...788...59W} and solved the dust radiative equilibrium temperature using the \textsc{torus} radiative transfer code \citep{2019A&C....27...63H}. We did this without self-consistently solving for the true hydrostatic structure (it was just imposed) and assumed ISM-like silicate grains. In the outer disc (beyond 20\,AU in a 100\,AU disc) we typically find that the ratio of the stellar irradiation floor temperature (equation \ref{equn:T_passive}) to the mid-plane dust radiative equilibrium temperature is $\approx 3-4$. We re-ran our MIST stellar luminosity SPH model grid, with the floor temperature attenuated by a factor 4 (but with a lower limit of 10\,K) and find that the resulting criterion is 28\,per cent of the way between the optically thick and thin limiting cases  }
\begin{equation}
    \frac{M_d}{M_*} < 0.17\left(\frac{R_d}{100\,\textrm{AU}}\right)^{1/2}\left(\frac{M_*}{M_\odot}\right)^{-1/2}. 
    \label{equn:irr_ratioRadeq}
\end{equation}
{As already mentioned, we have only briefly compared the passive and radiative equilibrium temperatures in a few cases so this should be treated with caution. Nevertheless it gives an indication of how irradiation will impact disc axisymmetry in reality compared to the limiting cases. } 

\section{Discussion}
\subsection{Implications for understanding the formation of Trappist-1 like systems}
The discovery of 7 roughly Earth mass planets in close orbit of the 0.08\,M$_\odot$ M dwarf Trappist-1 \citep{2017Natur.542..456G,  2017NatAs...1E.129L} begs for a theoretical understanding of the formation mechanism of such a treasure trove of planets. In addition to understanding the formation mechanism for this particular system, ultimately we need to understand how frequently such systems should be able to arise. This will likely be a function of the possible initial disc parameters -- which our models in this paper constrain (or, as it turns out, permit to vary somewhat dramatically) -- as well as of the environment in which the proto-Trappist-1 resides \citep{2018MNRAS.475.5460H}. After all, other low mass M dwarfs do not so far seem to reveal similar caches of planets.

\cite{2017A&A...604A...1O} developed a promising model for the formation of Trappist-1 in which grains drift in to the water snow line and there undergo planetesimal formation through the streaming instability. These planetesimals then migrate inwards and undergo extremely efficient pebble accretion \citep[this is contrasted with planetesimal accretion by][]{2019A&A...631A...7C}. Eventually they settle into mean motion resonance, with the innermost two planets not exhibiting this due to clearing of the inner disc. \cite{2017A&A...604A...1O} assumed $M_d/M_*=0.04$ for a 200\,AU disc, inferred a pebble accretion efficiency of about 25\,per cent and assumed that half of the drifting dust makes it to the water snow line. They also assumed that the initial dust-to-gas mass ratio (metallicity) was a factor two higher than canonical, i.e. $Z=0.02$. \cite{2019arXiv190600669S} recently followed up on that paradigm using a hybrid N-body, viscous evolution and analytic models. They found that indeed this process can lead to very efficient planet formation, converting more than half of the dust in their initial model into planets. 

A puzzle remaining from these models is why observationally we do not see all low mass M dwarfs with so many inner planets given the ease with which they are expected to form. \cite{2019arXiv190600669S} proposed that this might be due to dynamical instability of the orbital configurations . 

So we do have a candidate formation mechanism that is very efficient. However some fraction of the initial dust in any disc will 

a) be accreted onto the parent star, 

b) be stranded at larger radii in the disc at other snow lines (e.g. the CO and CO$_2$ snow lines) and 

c) can be stripped in an externally driven photoevaporative wind. 

Although the above proposed formation mechanism operates very early, so do these dust depletion/stranding mechanisms. The move towards understanding the prevalence (or lack thereof) of Trappist-1 like planetary systems requires understanding the viable initial conditions and possible barriers to the operation of the very promising \cite{2017A&A...604A...1O} model.
 
 \begin{table*}
    \centering
    \begin{tabular}{c|c|c|c|c|c|c|c|c|c|c|c}
    \hline
     System    &  Stellar Mass & Observed Radius & Observed $M_d/M_*$ & Optically thick & Optically thin & Equation 12  \\
     & $M_\odot$ & of spiral features & & $M_d/M_*$  criterion & $M_d/M_*$  criterion & $M_d/M_*$  criterion \\
    \hline
     IM Lup    & $\sim 1$ & 200\,AU & 0.1--0.17 & 0.17 & 0.42 & 0.24  \\
     Elias 2-27  & 0.5--0.6 & 200\,AU & 0.28 & 0.22--0.24 & 0.55--0.6 & 0.31--0.34 \\
     Wa Oph 6   & 0.5--0.6 & 100\,AU & Unknown & 0.155--0.170 & 0.39--0.42 & 0.22--0.24  \\
    \hline
    \end{tabular}
    \caption{{DSHARP discs with spirals, their disc-to-star mass ratios and the disc-to-star mass ratio for deviations from axisymmetry according to our criteria. ``Equation 12'' refers to our modified criterion that is motivated by our dust radiative equilibrium temperature calculations.} }
    \label{tab:DSHARPSPIRALS}
\end{table*}

Our models in this paper completely permit stable discs with the higher disc-to-star mass ratios that are expected to be required from photoevaporation models of proto-Trappist-1 by \cite{2018MNRAS.475.5460H}, even in the face of efficient planet formation. It would therefore be interesting to adapt the models of \cite{2019arXiv190600669S} to account for environmental dust stripping, compensate for this with higher initial disc masses and understand the impact that this has on their planet formation paradigm. For example, the growth of grains through the streaming instability in a higher mass disc may well operate in a very different manner to that expected in \cite{2017A&A...604A...1O} as discussed by \cite{2019MNRAS.485.4465F}.

In reality it is likely that a high efficiency formation mechanism in as low a disc mass as possible will be a favoured mechanism by which the Trappist-1 planets formed. However it is also worth retaining the idea that discs around low-mass stars can remain stable {against fragmentation} even if the disc mass is high.  Consequently, the formation process could actually be quite inefficient ($<10\%$). However any alternative mechanism utilising the possible higher disc mass would still have to explain the inferred water fraction of the Trappist-1 planets at $\sim10\%$ \citep{2018A&A...613A..68G, 2018ApJ...865...20D} which \cite{2019arXiv190600669S}  \citep[and also][]{2019A&A...631A...7C} successfully reproduces.

\subsection{Implication of GJ 3512}
{\cite{2019arXiv190912174M} recently detected a 0.46\,$M_{jup}$ planet around the 0.12\,$M_\odot$ star GJ 3512, which challenges typical models of planet formation around low-mass stars, since massive planets are thought to form via gravitational instability \citep[this notion was suggested by][for the case GJ 3512]{2019arXiv190912174M}. Although our general conclusion is that low-mass stars can indeed support non-fragmenting discs at high disc-to-star mass ratios, our results are not at odds with this discovery so long as the GI zone in the disc is optically thick (otherwise the required disc-to-star mass ratios get unrealistically high, at of order the stellar mass). If a 50\,AU disc is optically thick (akin to our non-irradiated case) then a disc-to-star mass ratio in the range 0.3--0.5 would exhibit the effects of self-gravity and become unstable towards the upper end. If it were GI that produced the massive planet around GJ 3512, the coupling with our results implies that discs around low-mass stars must be capable of supporting very high disc-to-star mass ratios ($M_d/M_*>0.3$) in reality.  }

\subsection{Implications for detecting non-axisymmetric self-gravitating discs}
A boundary to our full understanding of gravitational instability and its role in planet formation and disc evolution is that it has been difficult to identify observationally. This is because it is most likely to onset early, {likely when the system is embedded in the natal molecular cloud}, and may be active for only a small fraction of the disc lifetime,  making it relatively rare to catch in action. Our models introduce a further complication in that we expect it to also preferentially onset around higher mass stars (with the caveat that the initial disc-to-star mass ratio distribution as a function of stellar mass is unknown). GI is hence also more pervasive in the less common range of the initial mass function. We therefore anticipate that the best targets for identifying gravitational instability is discs around young stars of Solar mass and above. 

Given the above, a massive young disc like IM Lup \citep[disc-to-star mass ratio of $\sim0.17$ and a roughly Solar mass star][]{2016ApJ...832..110C} is hence a reasonable candidate and was recently shown to exhibit spirals in the DSHARP survey \citep{2018ApJ...869L..41A, 2018ApJ...869L..43H}. Though \cite{2016ApJ...832..110C} infer a minimum $Q$ of only 3.7 from their models, as \cite{2018ApJ...869L..43H} point out the estimation of disc masses is notoriously difficult and their observations are consistent with the idea that the IM Lup spirals result from GI.

The other DSHARP discs with spirals were WaOph 6 and Elias 2-27, the latter of which in particular has been the subject of much attention as a possible self-gravitating disc \citep[e.g.][]{2016Sci...353.1519P, 2017ApJ...835L..11T, 2017ApJ...839L..24M, 2018MNRAS.477.1004H, 2018ApJ...860L...5F}.  The stellar masses of these two systems are roughly Solar and 0.5--0.6\,M$_\odot$ respectively. The spiral features of IM Lup, WaOph 6 and Elias 2-27 extend out to around 200, 100 and 200\,AU respectively and the ages of all 3 systems are $<1\,$Myr. 

{In Table \ref{tab:DSHARPSPIRALS} we summarise the DSHARP discs with spirals and compare their measured disc-to-star mass ratios with our criteria for axisymmetry.  Currently inferred (but highly uncertain, and generally thought to be underestimated) disc-to-star mass ratios for IM Lup, Wa Oph 6 and Elias 2-27 are $0.1-0.17$ \citep[where we use a range with an upper limit set by the mass over the entire disc inferred by][]{2016ApJ...832..110C} , unknown (to the best of our knowledge) and 0.28 \citep{2009ApJ...700.1502A, 2010A&A...521A..66R} respectively.}

{Both IM Lup and Elias 2-27 are consistent with having self-gravity induced spiral features if the discs are optically thick. In the extreme case of the discs being optically thin to the stellar irradiation, none of the discs would be expected to deviate from axisymmetry due to self-gravity. In order for self-gravity to explain the spirals in the DSHARP discs, they either need to be very optically thick to stellar irradiation or have masses higher than currently observationally inferred (which is possible given the high uncertainty on this measurement). }

If our results were extrapolated to much higher stellar masses like proto OB stars, we would anticipate much more widespread instability. Observations of such discs are challenging owing to their scarcity, distance, short lifetimes and embedded nature. However, where resolved observations are finally becoming available we do indeed seem to be detecting evidence of gravitational instability and even fragmentation \citep{2018ApJ...869L..24I, 2019MNRAS.482.4673J}. As a simple illustration if we use the inferred stellar and disc masses from \cite{2018ApJ...869L..24I} of 34.2 and 5.8\,M$_\odot$ respectively, as well as the inferred disc size of 850\,AU in equation \ref{equn:irr_ratioMIST} we get a critical ratio for axisymmetry of 0.15 and observed ratio of 0.17 and so would expect the instability that they have observed. Of course there is a severe limitation in assuming that our criterion can be quantitatively extrapolated to such high stellar masses, and investigating this with detailed models will take place in future work \citep{2020MNRAS.492.5041C}. 

\subsection{A note on the time evolution of the stellar luminosity vs disc mass}
In our models a higher stellar luminosity more effectively reduces gravitational instability, owing to the enhanced thermal support that it provides against collapse. Referring back to Figures \ref{fig:luminosities} and \ref{fig:luminosityEvo}, which show the luminosities we assume, the stellar luminosity drops substantially between the very early evolution (0.5\,Myr) and ZAMS, particularly at the low stellar mass end. This raises the notion that for some discs in which the mass did not evolve appreciably, the stability of the disc might change over time as the stellar luminosity decreases. That is, if there were an initial warm, stable, high mass disc, we can ask if it will later fragment as the luminosity and hence temperature drops. For example a 50\,AU, 0.3\,M$_\odot$ disc around a 0.5\,M$_\odot$ star would be stable at 0.5\,Myr according to our models, but would develop spirals once the luminosity drops to the ZAMS level. 

Recall that in Figure \ref{fig:luminosityEvo} we showed the time evolution of the MIST luminosities for the stellar masses considered in our grid. The stellar luminosity actually  drops towards the main sequence value over a timescale that is longer than the average disc lifetime \citep[e.g.][]{2001ApJ...553L.153H, 2015A&A...576A..52R} even in the Solar mass case where it takes $\sim5$,Myr to reach the ZAMS (the mean disc lifetime is $\sim3$\,Myr). In the lowest stellar mass case it takes hundreds of Myr to reach the ZAMS, far beyond any observed disc lifetime.

Generally then, we anticipate that the disc mass evolution will be more rapid than the variation in stellar luminosity and so that the disc stability will not vary in time with the stellar luminosity.

\section{Summary and Conclusions}
We consider here the evolution of self-gravitating discs around young stars as a function of the star--disc parameters. We do so using three approaches. The first is 1D time-dependent models of discs evolving under the action of self gravity. These do not follow the nature of instability directly, but the effective viscosity gives a proxy for the stability, with $\alpha\sim0.1$ discs {expected to fragment}. Our second approach involves assessing the stable disc configurations from self-gravitating pseudo-viscous accretion models, again using a critical pseudo-viscous $\alpha$ of $\alpha \sim 0.1$. 

Our primary approach is to use smoothed particle hydrodynamics calculations to directly probe what disc-to-star mass ratios can be sustained around stars of different masses, without inducing spirals or fragmenting. We also assess the impact of disc size, and of accounting for passive stellar irradiation. 

We draw the following main conclusions from this work. \\

1. We derive approximate criteria for whether a disc is axisymmetric, exhibits spirals, or undergoes fragmentation due to self-gravity. We do this for {optically thin} discs irradiated by very young (0.5\,Myr) stellar luminosities appropriate for the time at which discs are most likely to be subject to the effects of self-gravity, and also consider the case of no irradiation (i.e. the {limiting} case of a disc that is highly optically thick to the central star radiation). A disc is expected to be axisymmetric for these two cases if the disc-to-star mass ratio satisfies
\begin{equation}
\boxed{
    \begin{array}{rclr}
    \frac{M_d}{M_*} & < & 0.3\left(\frac{R_d}{100\,\textrm{AU}}\right)^{1/2}\left(\frac{M_*}{M_\odot}\right)^{-1/2}, & \textrm{Optically\,\,thin} \\
    
    \frac{M_d}{M_*} & < & 0.12\left(\frac{R_d}{100\,\textrm{AU}}\right)^{1/2}\left(\frac{M_*}{M_\odot}\right)^{-1/2}, & \textrm{Optically\,\,thick}
\end{array}
}
\end{equation}
respectively. {Comparing the optically thin model approach with some preliminary dust radiative equilibrium calculations, we suggest that in reality the true criterion lies closer to the non-irradiated form with a normalisation $\approx0.17$.} These criteria provide a convenient, simple and relatively quick means of estimating limits on disc masses for models of real (non-)axisymmetric disc systems. In the past, typically $M_d/M_* < 0.1$ has been assumed, but this stems from constraints largely set by the Solar nebula. Note that we distinguish between axisymmetry and spirals with these criteria, rather than ``stability'' and ``instability'', since spirals can be exhibited in quasi-steady discs \citep{2001ApJ...553..174G, 2005MNRAS.364L..56R}.   \\

2. Applying our models to the DSHARP \citep{2018ApJ...869L..43H} discs with spirals and disc mass estimates, IM Lup and Elias 2-27, we find that {both would be consistent with self-gravity induced spirals if they are optically thick to the stellar irradiation, but neither are consistent with self-gravity induced spirals in the case of the disc being optically thin.}
 There is also the substantial caveat that disc masses are highly uncertain. For the other candidate with spirals identified in the DSHARP survey without a disc mass estimate, WaOph 6, we expect a required disc mass of $0.155-0.42$\,M$_\odot$ for self gravity induced spirals in the limits of being optically thick and optically thin. Optimistically extrapolating our criteria to proto O stars predicts disc spirals in the fragmenting system observed by \cite{2018ApJ...869L..24I} \\

3. Very low-mass stars ($\sim0.1\,$M$_\odot$) can sustain very high disc-to-star mass ratios before becoming {so gravitationally unstable that they undergo fragmentation.} This could play an important role in providing sufficient material to produce Trappist-1 like systems in the face of processes that deplete the dust reservoir (accretion, stranding at pressure maxima, external photoevaporation) even in the face of extremely efficient planet formation like that being proposed by \cite{2017A&A...604A...1O}. {Conversely, GI in discs around young stars is possible if they are optically thick, but still requires a disc-to star mass ratio > 0.3 for a 50\,AU disc.}  \\

\section*{Acknowledgements}
{We thank the reviewer for their thorough assessment of this paper, in particular regarding the methodology and robustness of results. }

TJH wishes to note that this work involved unusually strong input from all authors. We thank Ben Lewis and Matthew Bate for useful discussions regarding \textsc{sphNG}.  {We also thank Richard Booth for useful discussions.}

TJH and FM acknowledge support from the Royal Society Dorothy Hodgkin Fellowship. For part of this work TJH was funded by an Imperial College Junior Research Fellowship.  EA acknowledges an RAS summer studentship grant and the Imperial College UROP scheme.
CH is a Winton Fellow and this research has been supported by Winton Philanthropies / The David and Claudia Harding Foundation. CH has received funding from the European Union’s Horizon2020 research and innovation program under the Marie Sklodowska-Curie grant agreement No 823823 (RISE DUST-BUSTERS project). 

This work made use of the \textsc{splash} software \citep{2007PASA...24..159P}.

This work was performed using the DiRAC Data Intensive service at Leicester,
operated by the University of Leicester IT Services, which forms part of the
STFC DiRAC HPC Facility (www.dirac.ac.uk). The equipment was funded by BEIS
capital funding via STFC capital grants ST/K000373/1 and ST/R002363/1 and STFC
DiRAC Operations grant ST/R001014/1. DiRAC is part of the National
e-Infrastructure.

%%%%%%%%%%%%%%%%%%%%%%%%%%%%%%%%%%%%%%%%%%%%%%%%%%


\bibliographystyle{mnras}
\bibliography{references}

\appendix

\section{Resolution testing}
\label{sec:resConvTest}
\begin{figure}
    \includegraphics[width=8cm]{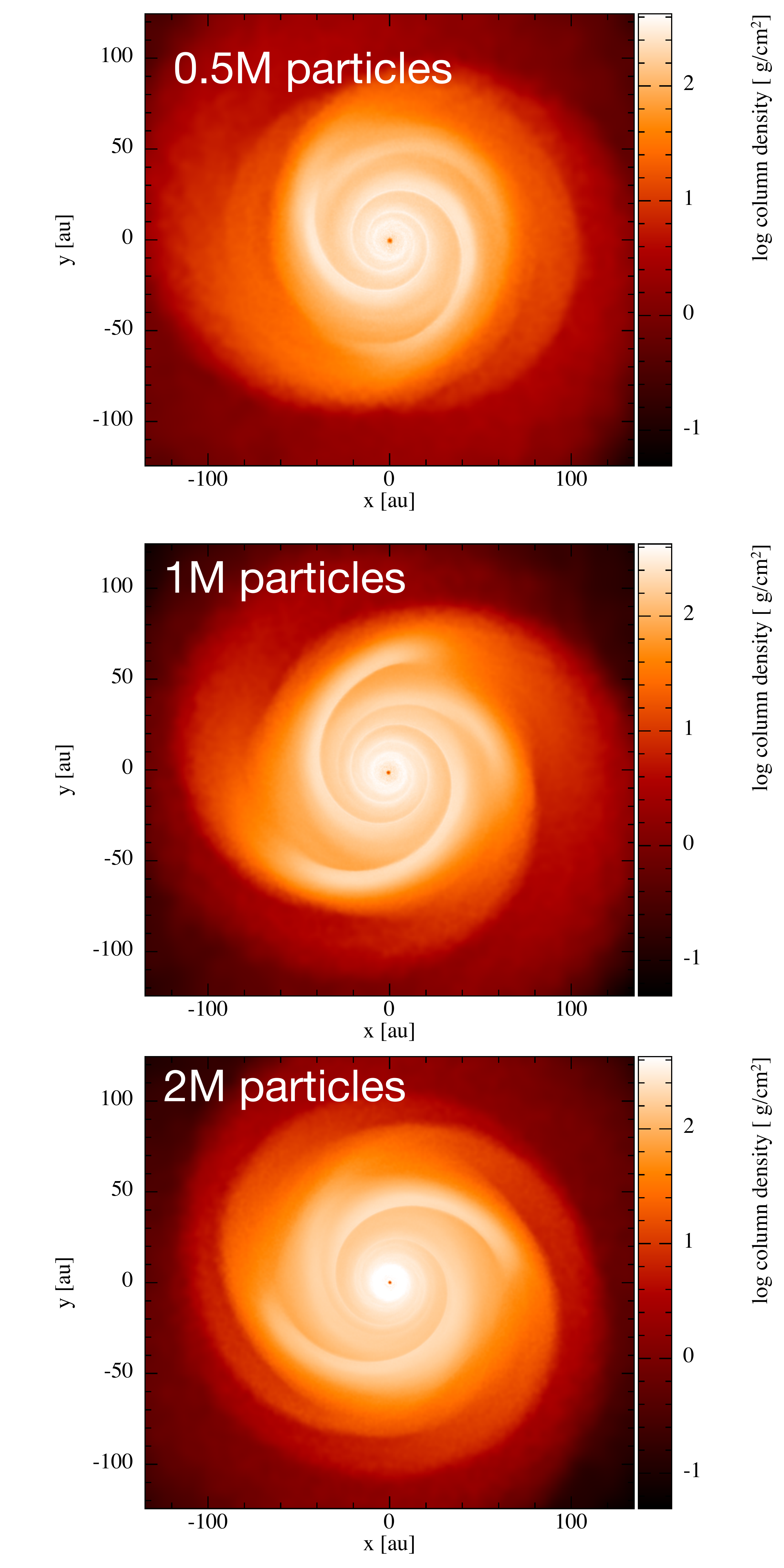}
    \caption{A resolution comparison of the MIST irradiation 1\,$M_\odot$ star model with a 100\,AU disc and disc-to-star mass ratio of 0.3.  The gross behaviour is the same in each case. }
    \label{fig:resolutionConv}
\end{figure}

\begin{figure}
    \centering
    \includegraphics[width=8cm]{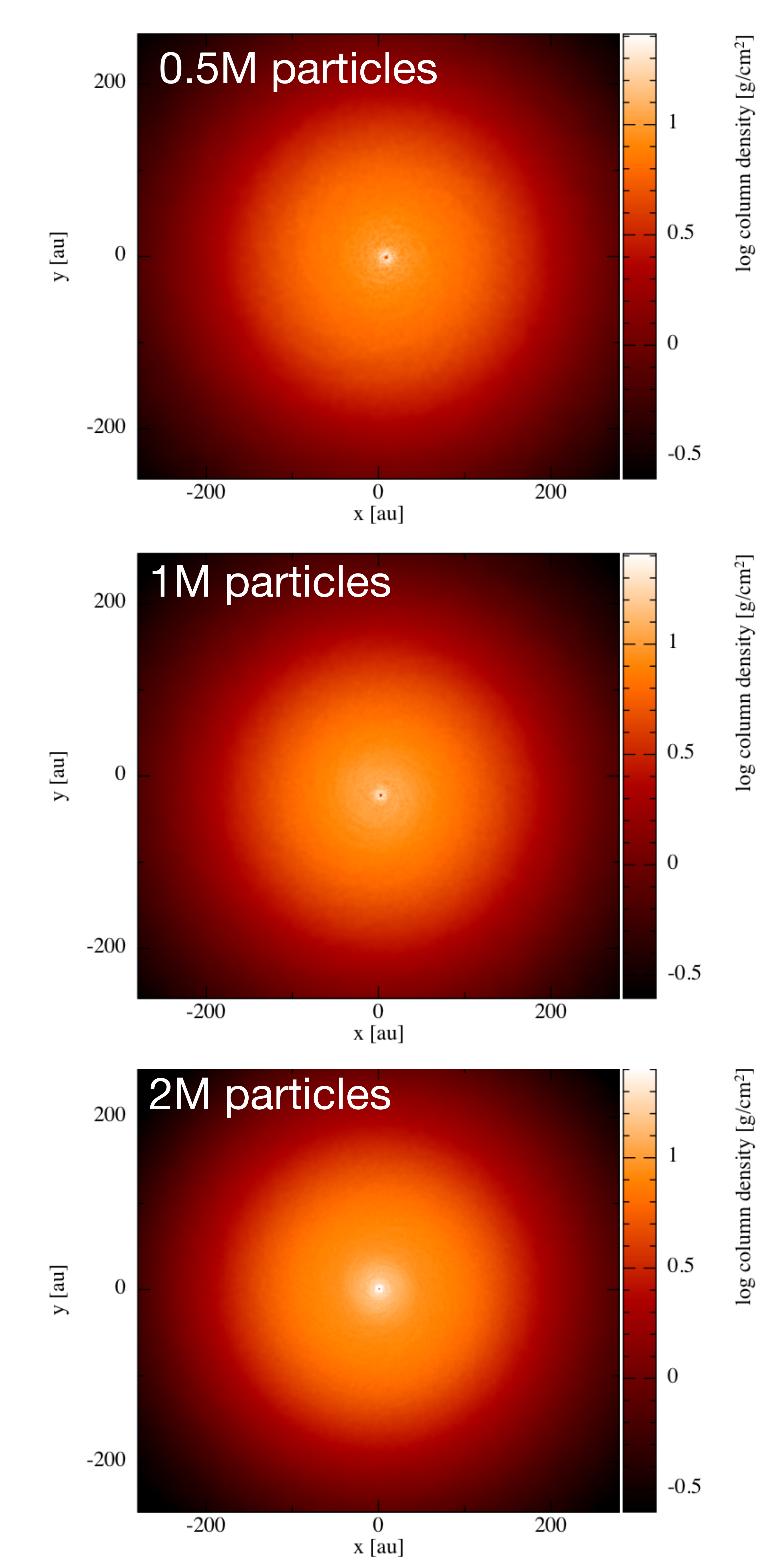}
    \caption{A resolution comparison of the MIST irradiation 0.1\,$M_\odot$ star model with a 200\,AU disc and disc-to-star mass ratio of 1.4. The gross behaviour is the same in each case. }
    \label{fig:resolutionConv2}
\end{figure}

\begin{figure}
    \includegraphics[width=8cm]{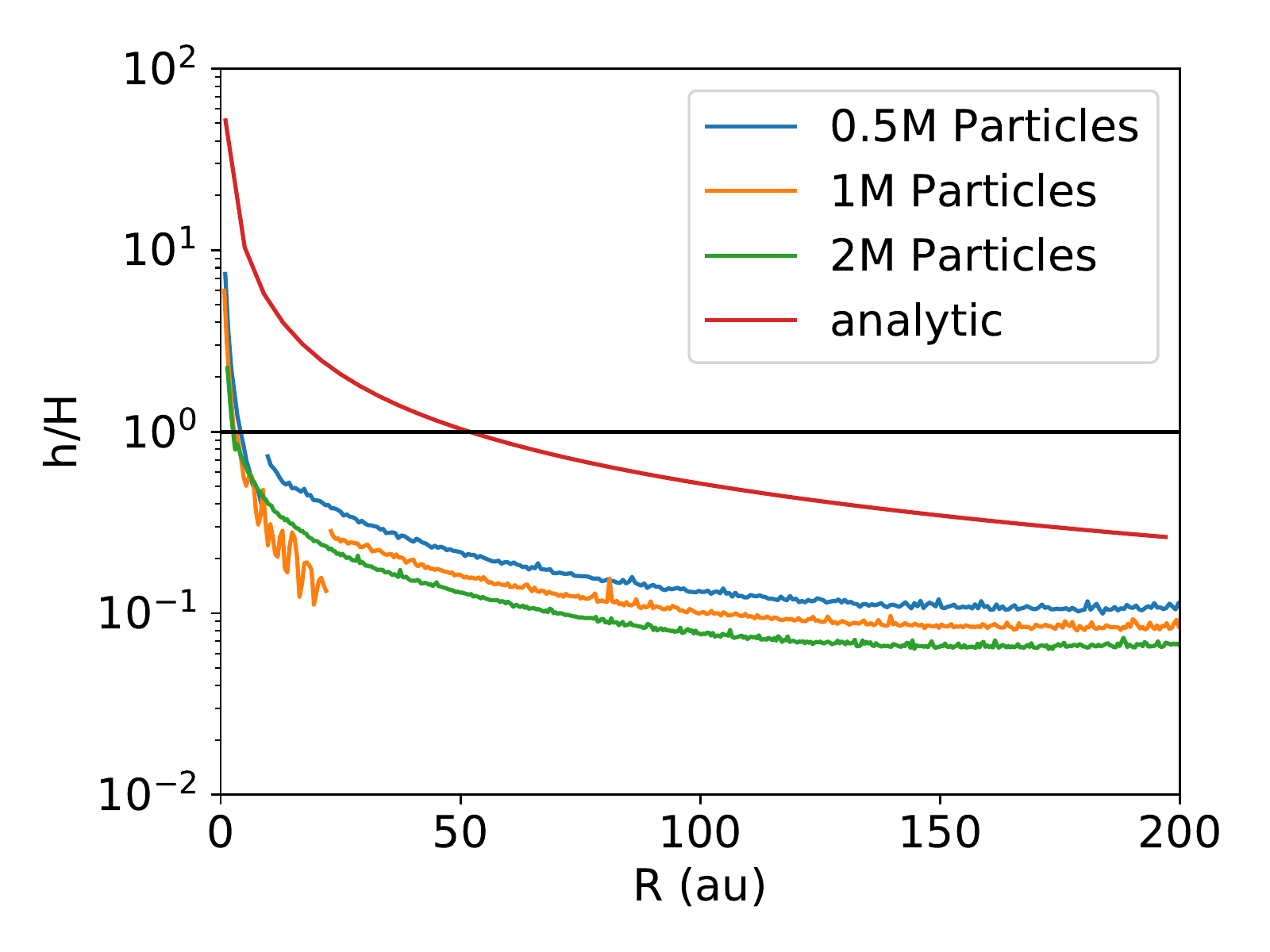}
    \caption{{The ratio of mean smoothing length to thermal scale height. The "analytic" line is akin to that in Figure \ref{fig:resolution}. The other lines are computed from the $0.1\,$M$_\odot$ star $M_d/M_*=1.4$ model with stellar irradiation that we used for convergence testing. An $h/H\sim0.25$ is required to capture self-gravitating discs. Our models that include heating from the central source are our best resolved (since the temperature in the disc is higher) and so the the ability of irradiated discs to support high disc-to-star mass ratios cannot be due to them being under-resolved. }}
    \label{fig:resolutionHeated}
\end{figure}

{As discussed in section \ref{sec:resolution}, to enable us to run a large grid our calculations use 0.5\,M particles each and are therefore not extremely well resolved. We hence re-ran the MIST 1\,M$_\odot$ star models using 1 and 2 million particles. The qualtitative behaviour is still the same, as summarised in Table 1. Snapshots of the $M_d/M_*=0.3$ case are shown in Figure \ref{fig:resolutionConv}. We also ran the extreme $0.1$\,M$_\odot$ star $M_d/M_*=1.4$ MIST model with 1 and 2 million particles, again verifying that its axisymmetry remains unchanged (see Figure \ref{fig:resolutionConv2}). We also note that the models with heating by the host star are better resolved (owing to the warmer temperatures and larger thermal scale height) which we illustrate in Figure \ref{fig:resolutionHeated}. The stability of irradiated discs with high disc-to-star mass ratios is therefore not due to the resolution. }

 \begin{table}
    \centering
    \begin{tabular}{c|c|c|c|c|}
    \hline
     $M_*$ & Mass ratio    &  0.5M particles & 1M particles & 2M particles \\
    \hline
    1 & 0.1  & Axisymmetric & Axisymmetric& Axisymmetric \\
    1 & 0.2  & Axisymmetric & Axisymmetric & Axisymmetric \\
    1 & 0.3  & Spirals & Spirals & Spirals \\
    1 & 0.4  & Spirals & Spirals & Spirals \\
    1 & 0.5  & Spirals & Spirals & Spirals \\
    0.1 & 1.4 &Axisymmetric &Axisymmetric &Axisymmetric \\
    \hline
    \end{tabular}
    \caption{{The outcomes of a sample of our models at different resolution} }
    \label{tab:resConvTest}
\end{table}

\section{Gallery of end states}
\begin{figure*}
    %\centering
    \vspace{-0.4cm}
    \includegraphics[width=20cm, angle=90]{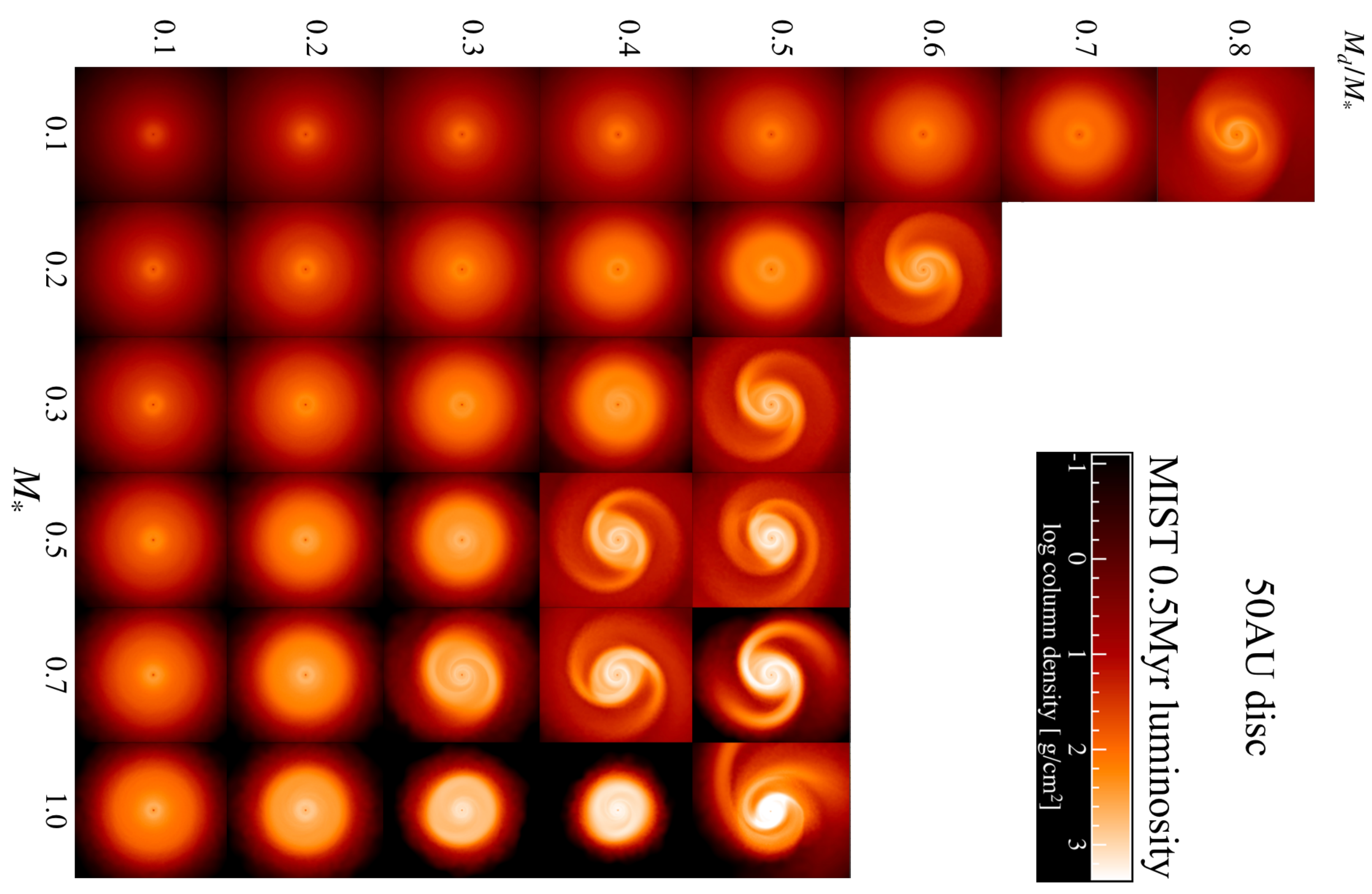}
    \vspace{-0.2cm}    
    \caption{Gallery of 50\,AU disc models with 0.5\,Myr MIST stellar luminosities. Shown is the surface density viewed face on. Columns are different stellar mass, rows different disc-to-star mass ratio. }
    \label{fig:gallery50}
\end{figure*}

\begin{figure*}
  %  \centering
    \vspace{-0.4cm}  
    \includegraphics[width=24cm, angle=90]{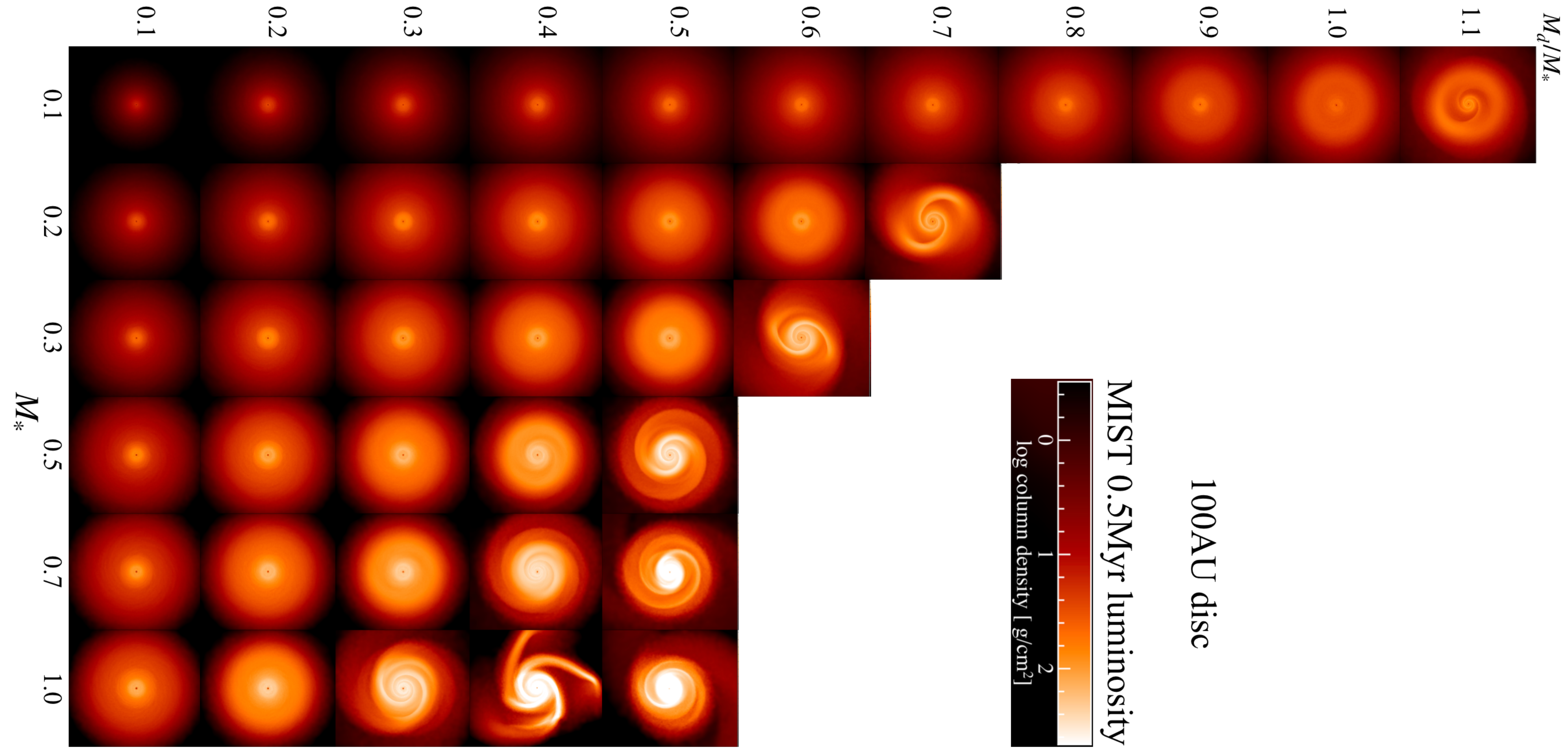}
    \vspace{-0.2cm}
    \caption{Gallery of 100\,AU disc models with 0.5\,Myr MIST stellar luminosities. Shown is the surface density viewed face on. Columns are different stellar mass, rows different disc-to-star mass ratio.  }
    \label{fig:gallery100}
\end{figure*}

\begin{figure*}
 %   \centering
    \vspace{-0.4cm}  
    \includegraphics[width=24cm, angle=90]{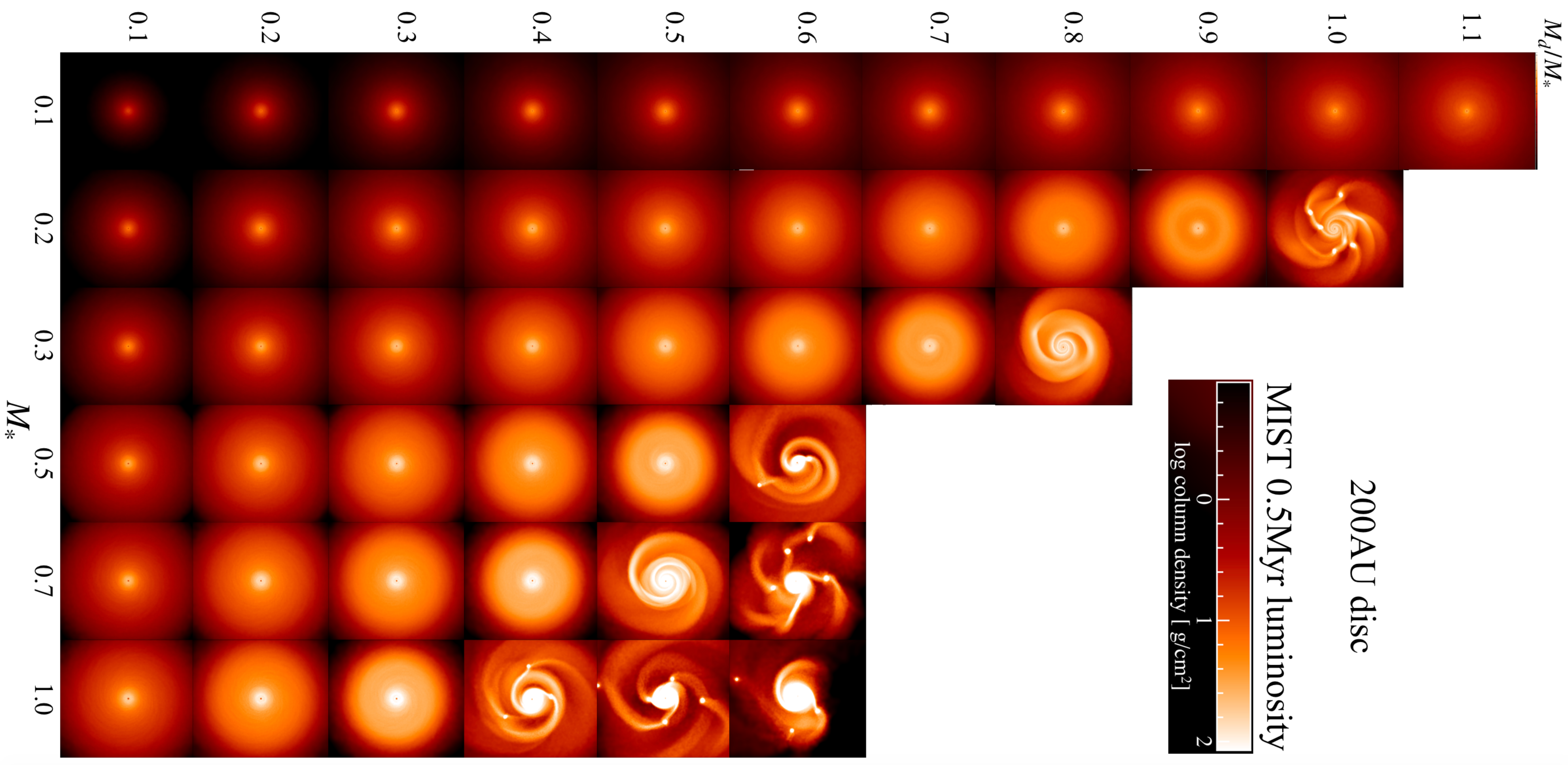}
    \vspace{-0.2cm}
    \caption{Gallery of 200\,AU disc models with 0.5\,Myr MIST stellar luminosities. Shown is the surface density viewed face on. Columns are different stellar mass, rows different disc-to-star mass ratio.  }
    \label{fig:gallery200}
\end{figure*}

In Figures \ref{fig:gallery50} through \ref{fig:gallery200} we present galleries of final snapshots from our {optically thin} models. These are the surface densities for the disc viewed face on.

% Don't change these lines
\bsp	% typesetting comment
\label{lastpage}
\end{document}